\documentclass[twocolumn,twocolappendix]{aastex63}

\usepackage{amsmath}
\usepackage{bbm,bm}
\usepackage{multirow}
\usepackage{colortbl}
\usepackage{preamble}
\usepackage{hyperref}
\usepackage{grffile}

\renewcommand{\log}{{\rm log_{\small 10}}}

\defcitealias{2010ApJ...721..975O}{OML10}
\defcitealias{2013ApJ...776....1K}{KOK13}
\defcitealias{2022ApJ...936..137O}{OK22}
\defcitealias{KGKO}{KGKO}
\defcitealias{2017ApJ...846..133K}{KO17}

\graphicspath{{./}{paper-figures/}{paper-tables/}}

\newcommand{\REV}[2]{{ #2}}
\shorttitle{Multiphase ISM and SFRs in TIGRESS-NCR}
\shortauthors{Kim et al.}

\begin{document}

\title{Introducing TIGRESS-NCR: I. Co-Regulation of the Multiphase Interstellar Medium and Star Formation Rates}

\author[0000-0003-2896-3725]{Chang-Goo Kim}
\affiliation{Department of Astrophysical Sciences, Princeton University, 4 Ivy Lane, Princeton, NJ 08544, USA}
\email{cgkim@astro.princeton.edu}

\author[0000-0001-6228-8634]{Jeong-Gyu Kim}
\affiliation{Division of Science, National Astronomical Observatory of Japan, Mitaka, Tokyo 181-0015, Japan}
\affiliation{Korea Astronomy and Space Science Institute, Daejeon 34055, Republic Of Korea}
\affiliation{Department of Astrophysical Sciences, Princeton University, 4 Ivy Lane, Princeton, NJ 08544, USA}

\author[0000-0003-1613-6263]{Munan Gong}
\affiliation{Max-Planck Institute for Extraterrestrial Physics, Garching near Munich, D-85748, Germany}

\author[0000-0002-0509-9113]{Eve C. Ostriker}
\affiliation{Department of Astrophysical Sciences, Princeton University, 4 Ivy Lane, Princeton, NJ 08544, USA}

\begin{abstract}
Massive, young stars are the main source of energy that maintains multiphase structure and turbulence in the interstellar medium (ISM), and without this ``feedback'' the star formation rate (SFR) would be much higher than is observed.
Rapid energy loss in the ISM and efficient energy recovery by stellar feedback lead to co-regulation of SFRs and the ISM state.
Realistic approaches to this problem should solve for the dynamical evolution of the ISM, including star formation, and the input of feedback energy self-consistently and accurately.
Here, we present the TIGRESS-NCR numerical framework, in which UV radiation, supernovae, cooling and heating processes, and gravitational collapse are modeled explicitly.
We use an adaptive ray tracing method for UV radiation transfer from star clusters represented by sink particles, accounting for attenuation by dust and gas.
We solve photon-driven chemical equations to determine the abundances of H (time-dependent) and C/O-bearing species (steady-state), which then set cooling and heating rates self-consistently.
Applying these methods, we present high-resolution magnetohydrodynamics simulations of differentially rotating local galactic disks representing typical conditions of nearby star-forming galaxies.
We analyze ISM properties and phase distributions and show good agreement with existing multiwavelength galactic observations.
We measure midplane pressure components (turbulent, thermal, and magnetic) and the weight, demonstrating that vertical dynamical equilibrium holds.
We quantify the ratios of pressure components to the SFR surface density, which we call the \emph{feedback yields}.
The TIGRESS-NCR framework will allow for a wide range of parameter exploration, including in low metallicity systems.
\end{abstract}
\keywords{Interstellar medium (847); Star formation (1569); Stellar feedback (1602); Magnetohydrodynamical simulations (1966); Radiative transfer simulations (1967)}

\section{Introduction}\label{sec:intro}

The interstellar medium (ISM) is at the core of the ecosystem in star-forming galaxies.
The ISM gives birth to stars and also processes the energy and metals these stars produce, using the majority in maintaining the ISM state while expelling a fraction to larger scales.
Modeling the ISM is fundamental to astrophysics, but challenging for many reasons.

Proper treatment of ISM dynamics and energetics must involve simultaneous modeling of the formation and evolution of massive young stars,
encompassing the physics that controls star formation, i.e., gravity, turbulence, and magnetic fields \citep[][]{2007ARA&A..45..565M}.
The ISM is highly dissipative, quickly losing energy through radiative processes \citep[e.g.,][]{1978ppim.book.....S,2011piim.book.....D}.
Without continuous and efficient energy inputs,  rapid gravitational collapse (approaching the free fall rate) would occur, which would produce far higher
star formation rates (SFRs) than those observed \citep[e.g.,][]{2022AJ....164...43S}.
Stars can provide the necessary energy, with UV radiation and supernovae (SNe) from massive young stars the two most energetically dominant channels \citep[e.g.,][]{1999ApJS..123....3L}.

UV radiation is the primary driver of key thermal and chemical processes in warm and cold ISM phases, setting cooling and heating rates \citep[][]{2022ARA&A..60..247W}.
The gas thermal pressure in warm and cold ISM phases
depends on the balance of heating and cooling \citep{1969ApJ...155L.149F,1995ApJ...443..152W}, and at a given temperature, the chemical state of the most abundant atom, hydrogen, can vary significantly: the warm medium ($T\sim10^4\Kel$) can be neutral or ionized, and the cold medium ($T\sim 10^2\Kel$) can be neutral or molecular. Thermal and chemical states depend on the strength of the UV radiation field as well as the cosmic ray (CR) rate \citep{2003ApJ...587..278W}, and these vary spatially due to proximity of sources and shielding by the highly inhomogeneous structure \citep[e.g.,][]{2017MNRAS.466.3293P}.
Additionally, shocks driven by SNe both accelerate and heat the gas \citep{1972ApJ...178..143C,1988ApJ...334..252C}. Given the galactic SN rate, the hot medium ($T\sim 10^6\Kel$) created by SN shocks can occupy a large volume \citep{1974ApJ...189L.105C,1977ApJ...218..148M} and break out the disk \citep{1976ApJ...205..762S}. Turbulence in the warm and cold ISM is driven by the interaction with expanding supernova-heated bubbles \citep[e.g.,][]{1999A&A...350..230K,2009ApJ...704..137J,2013ApJ...776....1K,2014A&A...570A..81H}.

From many decades of research, a consensus has been reached that multiple distinct phases coexist in the ISM, spanning a wide range of temperature and density \citep[e.g.,][]{2001RvMP...73.1031F,2005ARA&A..43..337C}.
Furthermore, because the ISM is dynamic, the thermal and chemical states of any given gas parcel are transient, and states that would traditionally be considered unstable are continuously repopulated.
Extensive observational surveys of the multiphase ISM have detailed the properties of the
cold molecular gas \citep{2001ApJ...547..792D, 2015ARA&A..53..583H}; the cold and warm atomic gas \citep{2016A&A...594A.116H,2018ApJS..234....2P}, with evidence of a significant fraction being in the unstable temperature regime \citep{2003ApJ...586.1067H,2018ApJS..238...14M}; the warm ionized medium \citep{2008ApJ...686..363H,2017ApJ...838...43K};  and the hot ionized medium \citep{1987ARA&A..25..303C,2008ApJS..176...59B}.

The existence of the multiphase ISM and the ubiquitous high-level turbulence in it are clear evidence that stellar feedback energy is effectively coupled to the ISM.
Feedback leads to inefficient star formation in terms of gas consumption, resulting in observed SFRs that are two orders of magnitude lower than the free-fall rate \citep{2018ApJ...861L..18U}.
Because energy dissipation leads to localized collapse and star formation, while the bulk ISM's energy loss can be efficiently recovered from stellar feedback,\footnote{
    \REV{}{There exist other sources of turbulent energy (which can then feed thermal energy through dissipation) including
    large scale gravitational instabilities
    \citep[e.g.,][]{1965MNRAS.130..125G,1981ApJ...246L.151F,2007ApJ...660.1232K,2010MNRAS.409.1088B,2018MNRAS.477.2716K,2018ApJ...854..100M,2020ApJ...896L..34B}
    and magnetorotational instabilities \citep[e.g.,][]{1999ApJ...511..660S,2003ApJ...599.1157K,2005ApJ...629..849P,2007ApJ...663..183P}.
    However, it seems unlikely that gravitational instability alone can provide sufficient turbulent support to control star formation, given the unrealistically high star formation rates in simulations with no feedback \citep{2011MNRAS.417..950H,2013ApJ...770...25A}.
    At least in nearby normal star-forming galaxies, it appears that feedback is sufficient to drive observed turbulence
    \citep[e.g.][]{2020A&A...641A..70B}, although it is possible that other sources of turbulence become more important under more extreme conditions.}}
the ISM's physical state and the SFR are intimately connected and in fact must be
 \emph{co-regulated} \citep[][]{2010ApJ...721..975O}.

To quantitatively understand the co-regulation process in the star-forming ISM, a holistic approach is required, using direct numerical simulations to solve the magnetohydrodynamics (MHD) equations with gravity and explicit cooling and heating.
These numerical simulations explicitly model the ISM in all phases, while tracking star formation in gravitationally collapsing regions and directly following energy inputs from recently formed stars. In order to be self-consistent, all gas heating and turbulence driving must either originate in energy inputs from stars, or develop as a consequence of naturally-occurring ISM dynamics (driven by gravity, shear, etc). Realistic numerical simulations of the star-forming multiphase ISM require both high mass and spatial resolution to resolve both the mass-dominating (cold and warm) and volume-filling (warm and hot) components.
Given the strict resolution requirements for multiphase ISM simulations, to date, the outer dimensions of simulation domains are still limited to a few kpc, corresponding either to low-mass dwarf galaxies \citep{2017MNRAS.471.2151H,2022arXiv220509774S} or to portions of larger galactic disks, using vertically-stratified boxes \citep{2017ApJ...846..133K,2017MNRAS.466.1903G,2017MNRAS.466.3293P,2021MNRAS.504.1039R,2020MNRAS.491.2088K}.

The TIGRESS framework was developed by \citet{2017ApJ...846..133K} to study the star-forming multiphase ISM, including a full complement of physics modules, sufficient resolution to follow key processes, and computational performance that enables both long-term evolution and comparative study of multiple galactic environments.
In the TIGRESS framework, the  MHD equations in a local patch of a differentially rotating galactic disk are solved with the grid-based code {\tt Athena} \citep{2008ApJS..178..137S,2009NewA...14..139S}. Self-gravity of gas is included, together with a fixed vertical potential from stars and dark matter. Cooling is modeled by a temperature-dependent tabulated cooling function appropriate for solar metallicity \citep{1993ApJS...88..253S,2002ApJ...564L..97K}. Sink particles representing star clusters are introduced within cells undergoing runaway gravitational collapse  \citep{2013ApJS..204....8G}. Photoelectric heating by FUV radiation is set to scale with the globally attenuated FUV luminosity from star clusters formed in the simulation. Explosions of individual SNe are directly modeled,
resolving the Sedov-Taylor stage during which the radial momentum of expanding bubbles is built up and the hot ISM is created in strong shocks \citep{2015ApJ...802...99K}.
Systems modeled by the TIGRESS framework successfully evolve to a quasi-steady state over many star formation and feedback cycles. A large suite of individual TIGRESS simulations covering varying galactic conditions shows that in all cases a state with self-consistently regulated SFR and ISM state is reached \citep{2022ApJ...936..137O}, with multiphase outflows launched from the disk \citep{2018ApJ...853..173K,2020ApJ...894...12V,2020ApJ...900...61K}.
These TIGRESS simulations have also been used to study
cloud and star formation correlations \citep{2020ApJ...898...52M}, molecular chemistry \citep{2018ApJ...858...16G,2020ApJ...903..142G}, diffuse ionized gas \citep{2020ApJ...897..143K}, polarized dust emission \citep{2019ApJ...880..106K}, and CR transport \citep{2021ApJ...922...11A,2022ApJ...929..170A}.  In addition, the TIGRESS computational framework of \citet{2017ApJ...846..133K} has been extended to simulate regions with strong spiral structure \citep{2020ApJ...898...35K},  nuclear rings where bar-driven gas inflows accumulate \citep{2021ApJ...914....9M,2022ApJ...925...99M}, and ram pressure stripping by the intracluster medium \citep{2022ApJ...936..133C}.

This paper presents the first application of an extension of the TIGRESS framework, called TIGRESS-NCR, where ``NCR'' stands for Non-equilibrium Cooling and Radiation. \REV{TIGRESS-NCR includes}{The two salient new features of TIGRESS-NCR are} explicit UV radiation transfer using the adaptive ray tracing method implemented in {\tt Athena} \citep{2017ApJ...851...93K} and the photochemistry model introduced by \citet{KGKO}. We solve time-dependent chemical rate equations for hydrogen species and obtain other abundances assuming formation-destruction balance given the hydrogen species abundances. Cooling in the cold and warm medium  in TIGRESS-NCR is determined by abundances of hydrogen species (H, $\Hp$, $\HH$) and major coolants ($\mathrm{C^+}$, C, CO, O, $\mathrm{O^+}$). Cooling in warm ionized gas is treated with a nebular cooling function that assumes a fixed abundance pattern characteristic of photoionized regions.
High-temperature He and metal cooling assume collisional ionization equilibrium (CIE). To follow UV radiation, photon packets emanating from star clusters are transferred along rays through the ISM, with absorption by dust and gas. The major radiative heating channels (photoelectric and photoionization heating) and expansion of overpressurized \ion{H}{2} regions (driven by photoionized gas and radiation pressure) are directly modeled.
We also include cosmic-ray induced ionization and heating with an attenuation factor inversely proportional to an effective mean column density. Our new framework with adaptive ray tracing
improves upon the accuracy of radiation transfer solutions compared to other ISM simulations that adopt more  approximate methods, including the local attenuation and local ionization approach in \citet{2021ApJ...920...44H}, the tree-based backward radiation transfer method in \citet{2021MNRAS.504.1039R}, and the two-moment approach with M1 closure in \citet{2020MNRAS.491.2088K,2022arXiv221104626K}.

In this paper, we focus on technical aspects of the TIGRESS-NCR implementation, and demonstrate how the ISM state and SFR are co-regulated by the full physics treatments in the TIGRESS-NCR framework.
We consider two different galactic conditions for the models in this paper, one similar to the solar neighborhood, and one representing inner-galaxy environments.
In a companion paper, we use the TIGRESS-NCR implementation to conduct a set of controlled numerical experiments in which we turn on and off individual feedback channels and the magnetic field, in order to investigate the role of each physical process in regulating SFRs and ISM properties. In subsequent papers, we will present detailed analyses of radiation properties, ISM energetics, and galactic outflows.

We describe numerical details in \autoref{sec:methods}, drawing from \citet{2017ApJ...846..133K} and from \citet{KGKO}. TIGRESS-NCR specific treatments regarding truncation of rays for efficient calculations are detailed in \autoref{sec:rt} and \autoref{sec:A-rt}.
\autoref{sec:simulations} describes the ISM properties, energetics, and phase distributions for the two simulated galactic conditions. \REV{}{
New features of models enabled by the TIGRESS-NCR framework include maps of radiation fields and chemical abundances,  as well as a full phase separation using both temperature and hydrogen abundances.} \autoref{sec:prfm} examines SFRs and the ISM state in the context of the pressure-regulated, feedback-modulated (PRFM) star formation model \citep{2022ApJ...936..137O} by analyzing the midplane pressure components and their ratio to SFR surface density (feedback yields). \REV{}{ } \autoref{sec:summary_and_discussion} summarizes our simulation results and discusses observational constraints, also situating our work within the context of recent star-forming ISM numerical studies.

\section{Methods}\label{sec:methods}

In this section, we introduce the TIGRESS-NCR numerical framework.
This is an extension of the original TIGRESS framework \citep[][which we refer to as TIGRESS-classic hereafter]{2017ApJ...846..133K}  coupled with photochemistry and UV radiation transfer modules, as detailed in \citet{KGKO}.
We begin by describing the governing equations (\autoref{sec:eqn}), and then briefly summarize the methods for treating star formation and SNe (\autoref{sec:sfsn}), radiation transfer (\autoref{sec:rt}), and photochemistry and cooling/heating (\autoref{sec:pchem}).
\REV{}{Readers who are familiar with TIGRESS-classic can skip to the latter two sections for the new features.}

\subsection{Governing Equations}\label{sec:eqn}
We solve the MHD equations in a local Cartesian rotating frame, with background galactic differential rotation treated via the so-called shearing-box approximation \citep{1965MNRAS.130..125G,HGB1995}. We use the grid-based code
{\tt Athena} to solve the ideal MHD equations \citep{2008ApJS..178..137S,2009NewA...14..139S}, employing a high-order Godunov method combined with the constrained transport algorithm \citep{2008JCoPh.227.4123G}.

The conservation equations for gas mass, momentum, and total energy are, respectively,
\begin{equation}\label{eq:mass_con}
    \pderiv{\rho}{t} + \divergence{\rho\vel} = 0,
\end{equation}
\begin{align}\label{eq:mom_con}
    \pderiv{(\rho\vel)}{t} +\divergence[\sbrackets]{\rho \vel\vel + \rbrackets{P+P_B}\mathbb{I} - \frac{\Bvec\Bvec}{4\pi}} =
    \nonumber\\
 - \rho \nabla \Phi  - 2\Om\times   (\rho\vel) +\frad,
\end{align}
and
\begin{align}\label{eq:energy_con}
    \pderiv{\etot}{t} +
    \divergence[\sbrackets]{\vel(\etot+P+P_B)- \frac{\Bvec(\Bvec\cdot\vel)}{4\pi}} =
    \nonumber\\ \mathcal{G} - \mathcal{L} - (\rho\vel) \cdot\nabla{\Phi} +\vel\cdot\frad.
\end{align}
The magnetic field evolution is governed by the induction equation without explicit resistivity (ideal MHD):
\begin{equation}\label{eq:induction}
    \pderiv{\Bvec}{t}=\curl{\vel\times\Bvec},
\end{equation}
while $\Bvec$ must obey the divergence-free constraint
\begin{equation}\label{eq:divzero}
\nabla\cdot \Bvec =0.
\end{equation}
In the above, $\rho = \muH m_{\rm H}\nH$ is the gas density, $\nH$ the  number density of hydrogen nuclei, $\muH$  the mean molecular weight per H nucleus, and
$m_{\rm H}$  the mass of a hydrogen atom;
$\vel$ and $\Bvec$ are velocity and magnetic field vectors, respectively;
$P$ and $P_B=\Bvec\cdot\Bvec/(8\pi)$ are thermal and magnetic pressure, respectively;
$\etot=\rho v^2/2 + P/(\gamma -1) + P_B$ is the total energy density, where $\gamma=5/3$ is the adiabatic index.

We explicitly follow non-equilibrium abundances of molecular ($\xHH$) and ionized ($\xHII$) hydrogen by solving the transport of abundances with source terms,
\begin{equation}\label{eq:scalar}
  \pderiv{\rho_s}{t} + \nabla \cdot (\rho_s \vel) = \rho\mathcal{C}_s,
\end{equation}
where
$\rho_s = \rho x_s$ is the mass density of species $s$ ($\HH$ or $\Hp$), and $\mathcal{C}_s$ is the net creation rate coefficient.

On the right hand side of the momentum equation (\autoref{eq:mom_con}), we have source terms due to the total gravitational force ($-\rho \nabla \Phi$), Coriolis force ($-2\rho\vel\times\Om$), and radiation force ($\frad$) per unit volume. The total gravitational potential $\Phi = \Phi_{\rm sg} + \Phi_{\rm ext}(z) + \Phi_{\rm tidal}(x)$ includes the self-gravitational potential obtained as the solution of Poisson's equation (including contributions from both gas and young star clusters, represented numerically as sink/star particles),
\begin{equation}\label{eq:poisson}
    \nabla^2 \Phi_{\rm sg} = 4\pi G (\rho + \rho_{\rm sp}),
\end{equation}
the external gravitational potential in the vertical direction (fixed in time; see \citet{2017ApJ...846..133K} for the exact form), and the tidal potential which gives rise to the differential rotation of the background flow (non-rigid body rotation); see below for the last. In the energy equation (\autoref{eq:energy_con}), we then have mechanical energy source terms associated with the gravity and radiation pressure forces (there is no work from Coriolis forces) in addition to the
radiative heating and cooling terms ($\mathcal{G} - \mathcal{L}$).

We solve \autoref{eq:poisson} using a Fast Fourier Transform
method with shearing-periodic boundary conditions in the horizontal directions \citep{2001ApJ...553..174G} and open boundary conditions in the vertical direction \citep{2009ApJ...693.1316K}.
We include newly formed stars' gravity using the particle mesh method by depositing the particle mass using a triangle-shaped cloud to obtain $\rho_{\rm sp}$ \citep{2013ApJS..204....8G}.
The center of our domain corotates  with the  background galactic rotation speed at galactocentric radius $R_0$, i.e., $\Om=\Omega_0\zhat$, and we assume the galactic rotation curve is flat, i.e., the shear parameter $q\equiv - d\ln \Omega/d\ln R =1$. As a result, $\Phi_{\rm tidal}(x) = - q \Omega_0^2 x^2$,
where $x$ is the local radial coordinate ($x=0$ at the domain center). The source terms due to galactic differential rotation are included in the hyperbolic integrator by a semi-implicit method (Crank-Nicholson time differencing) as described by \citet{2010ApJS..189..142S}. The gravity source term is also included in the integrator,
while the radiation force and cooling/heating source terms are included using an operator-split approach (see below).

The main new features
in the TIGRESS-NCR framework
are the explicit treatments of chemical processes and associated cooling and heating terms. This is in contrast to the TIGRESS-classic framework, in which the heating rate per hydrogen $\Gamma$ is spatially constant (but time-variable), set via a simple scaling relative to the solar neighborhood value of the  globally-attenuated instantaneous FUV radiation field as produced by star cluster particles
 \citep{2020ApJ...900...61K}. In TIGRESS-classic, the cooling function $\Lambda(T)$ is only a function of temperature with a temperature dependent mean molecular weight $\mu(T)$ combining \citet{2002ApJ...564L..97K} and \citet{1993ApJS...88..253S}.

In this work, we  directly calculate chemical reaction rates and cooling/heating rates from key microphysical processes that depend on gas properties ($\nH$, $T$, $\vel$), species abundances ($x_s$), radiation fields in three UV bands ($\erad$; see \autoref{sec:rt}), and the CR ionization rate ($\xicr$).
Explicitly, we have
\begin{align}
    \mathcal{C}_s & \equiv \mathcal{C}_s(\nH, T, x_s, \Zd, \Zg,  \erad, \xicr) \,, \label{eq:Chem_def}
  \\
  \mathcal{G} & \equiv \nH\Gamma(\nH, T, x_s, \Zd, \Zg,  \erad, \xicr) \,, \label{eq:Gamma_def}
  \\
  \mathcal{L} & \equiv \nH^2\Lambda(\nH,T, x_s, \Zd, \Zg, \erad, |dv/dr|) \,.
  \label{eq:Lambda_def}
\end{align}
where $\Zg$ and $\Zd$ are the gas metallicity and dust abundance relative to solar neighborhood values.
Details of these functions are provided in \citet{KGKO}.
This paper assumes $\Zg = \Zd = 1$, corresponding to solar metallicity with abundances of \citet{Asplund09} and fractional mass of metals
$Z_{\rm g,\odot} = 0.014$; and mass of grain material relative to gas $0.0081$ \citep{Weingartner01a}. Although here we adopt globally constant values for $\Zg$ and $\Zd$, in principle they can be determined locally (with appropriate treatments for metal enrichment and dust formation and destruction processes), 
and the TIGRESS-NCR framework is applicable for a wide range of $\Zg$ and $\Zd$ except for gas with very low metal and dust content in the early universe.\footnote{For example, our model for the CO abundance has been tested limited ranges of parameter values \citep{2017ApJ...843...38G}. Also, our model does not include gas-phase H$_2$ formation and HD cooling, which can be important for very low-metallicity, dense gas \citep[e.g.,][]{2004ApJ...611...40C, 2005ApJ...626..627O}.}

As detailed in \citet{KGKO}, we note that the heating and cooling functions that we adopt follow \citet{1995ApJ...443..152W,2003ApJ...587..278W} in all essential aspects and produce results consistent with theirs for solar neighborhood conditions, while also allowing for varying metal and dust abundances as well as UV and CR fluxes \citep[see also][]{2019ApJ...881..160B}.
Because our simulations are time-dependent, they also make it possible to test the extent to which the predicted thermal equilibrium state is actually reached.

\subsection{Star Formation and Supernovae}\label{sec:sfsn}

In addition to the source terms given on the right hand sides of Equations~\eqref{eq:mass_con}--\eqref{eq:energy_con}, we also include sink and source \REV{and}{} terms associated with star formation and SN feedback.
The treatments of star formation and SN feedback using sink/star particles are identical to methods adopted for TIGRESS-classic.

We form and grow star cluster particles based on the sink particle treatment in {\tt Athena} first introduced by \citet{2013ApJS..204....8G} and modified further for the TIGRESS-classic framework \citep{2020ApJ...900...61K}.
Within the control volume ($3^3$ cells) surrounding a cell undergoing unresolved gravitational collapse, we create a sink particle by turning gas mass and momentum into a particle's mass and velocity. The collapse criteria are: (1) a cell's density is higher than a threshold Larson-Penston density depending on sound speed and numerical resolution, (2) the cell is at a local gravitational potential minimum, and (3) flows along all three Cartesian directions converge toward the cell.
Each particle represents a star cluster (with typical mass $>10^3\Msun$ for our adopted resolution) consisting of coeval stars from a fully-sampled initial mass function (IMF). We use the STARBURST99 stellar population synthesis (SPS) model \citep{1999ApJS..123....3L, 2014ApJS..212...14L} to determine SN rates for each star cluster, assuming a \citet{2001MNRAS.322..231K} IMF and Geneva evolutionary tracks for non-rotating stars.

Each star cluster hosts multiple SNe over its lifetime ($t_{\rm age}<40\Myr$). We assume 50\% of SN events are in binary OB systems, and if
an event was from a binary we inject a massless particle with a velocity kick \citep{2011MNRAS.414.3501E}. These runaway stars can travel far from the cluster particle before they explode as SNe. However, we do not consider runaways as sources of UV radiation because the computational cost of ray tracing would become too expensive if runaway sources were included, and tests show that they do not contribute significantly to the total luminosity or ionization budget \citep{2020ApJ...897..143K}.

For each SN event, we first calculate the enclosed mass $M_{\rm SNR}$ (sum of ejecta, $M_{\rm ej}=10\Msun$, and pre-existing ambient ISM) and mean density $n_{\rm amb}$ of the surrounding ISM within a spherical region with a radius of $3\Delta x$. If the calculated gas mass is less than the shell formation mass $M_{\rm sf} = 1540\Msun(n_{\rm amb}/\pcc)^{-0.33}$ when a remnant becomes radiative \citep{2015ApJ...802...99K}, a total of $E_{\rm SN}=10^{51}\erg$ energy is injected with a thermal to kinetic energy ratio consistent with that of the Sedov-Taylor stage of evolution, after averaging out the feedback injection region. If the Sedov-Taylor stage is unresolved (i.e., $M_{\rm SNR}>M_{\rm sf}$), a total of $p_{\rm SNR}=2.8\times10^5\Msun\kms(n_{\rm amb}/\pcc)^{-0.17}$ radial momentum is injected. Given the high resolution and natural clustering of SNe realized in our simulations, only a small fraction of SN events ($<10\%$) are realized in the form of pure momentum injection.

\subsection{UV Radiation Transfer and Cosmic Rays}\label{sec:rt}

All star cluster particles with (mass-weighted) age $t_{\rm age} \le 20 \Myr$ act as sources of UV radiation. Appendix C in \citet{KGKO} provides details regarding radiation characteristics of the adopted SPS model.

We divide UV radiation into three frequency bins: photoelectric (PE; $110.8\nm < \lambda < 206.6\nm$), Lyman-Werner (LW; $91.2\nm < \lambda < 110.8\nm$), and Lyman continuum (LyC; $\lambda < 91.2\nm$). Both LW and PE photons (collectively referred to as FUV) provide an important source of gas heating via the photoelectric effect when absorbed by small dust grains and large molecules. All FUV photons are attenuated by dust along rays, and the LW band photons also dissociate H$_2$ and CO, and ionize C. To compute the dissociating radiation field for H$_2$, we apply the \citet{Draine96} self-shielding function to the LW band, using the H$_2$ column density integrated from each source to each cell. The LyC photons (also referred to as EUV) ionize neutral hydrogen (H and H$_2$) and are attenuated by dust.

To follow the propagation of UV photons from young star clusters, we utilize the adaptive ray tracing module in the {\tt Athena} code \citep{2017ApJ...851...93K}.
After the hyperbolic part of the governing equations (including shearing-box and gravitational source terms) is integrated, we solve the time-independent UV radiation transfer equation,
\begin{equation} \label{eq:RT}
  \khat \cdot \nabla I_j = -\alpha_j I_j
\end{equation}
for each frequency bin $j$ along a set of rays. Here, $I_j$ is the intensity, $\khat$ is the unit vector specifying the direction of radiation propagation, and $\alpha_j$ is the absorption cross section per unit volume. In brief,
$12 \times 4^4$ photon packets are injected for each band at the location of each source \REV{}{on a set of rays covering solid angles,} corresponding to HEALPix level 4 \citep{Gorski05}. Photon packets propagate radially outward, and rays are split into four children when needed to ensure that each cell is intersected by at least 4 rays per source. The optical depth of each ray through each intersecting cell is computed and used to apply the corresponding rate of energy and momentum deposition.
The radiation energy density ${\erad}_{,j}$ and flux ${\Frad}_{,j}$ in each cell are obtained by summing over contributions from all rays passing through it. We then have $\frad =\sum_j \rho\kappa_j{\Frad}_{,j}/c$, which we use to update the gas momentum and corresponding kinetic energy density; \REV{}{the values of ${\erad}_{,j}$ in each cell are employed in photochemistry and heating calculations (\autoref{sec:pchem}).}

As with other fluid properties, the shearing-periodic boundary conditions are implemented for the ray tracing. Photon packets crossing the local-radial ($\xhat$) edges of the computational domain are offset by the distance the boundaries have sheared in the local-azimuthal ($\yhat$) direction, and the position of sources is offset accordingly.
The boundary condition in the $y$ direction is periodic.

We terminate a ray if a photon packet exits the $z$-boundary of the computational domain or the optical depth measured from the source is larger than $\tau_{\rm max}=30$ in all frequency bins.
With just these basic ray termination conditions, however, we find that the computational cost of performing adaptive ray tracing every timestep is prohibitive.
To reduce the cost of ray tracing without losing accuracy too much, we adopt three modifications. (1) We perform ray tracing at intervals $\Delta t_{\rm 2p}$
based on the Courant–Friedrichs–Lewy (CFL) time step for the cold and warm gas at $T < 2.0 \times 10^4 \Kel$,
or whenever a new radiation source is created, or whenever an existing radiation source is removed.
(2) We put a hard limit on the maximum horizontal distance a ray may propagate from each source, which we denote as $\dxymax$.
(3) We terminate a ray in the FUV band if the ratio between the luminosity of the photon packet and the total luminosity of all sources in the computational domain falls below a small number $\epp$.
The first condition, on the interval for radiation updates, is justified because the hot gas has very low opacity and its interaction with radiation is minimal.
If there is no limitation on the maximum horizontal propagation distance of rays, the cost of ray tracing can become prohibitively high. We have found that imposing condition (2) reduces the cost to an acceptable level without significantly affecting the radiation field solution in the midplane region, provided $\dxymax$ is large enough (see \autoref{sec:A-rt-conv}).
The condition (3) limits the maximum distance traveled by photon packets originating from faint sources, without seriously degrading the accuracy of the radiation field.

Terminating rays based on $\dxymax$ and $\epp$ causes underestimation of the FUV radiation field at high altitudes. To address this issue without incurring  additional computational cost, we apply an analytic solution based on the plane-parallel radiation transfer (see \autoref{sec:A-rt-planeparallel}). We stop ray-tracing for the PE and LW bands at $|z| > \zpp$ and measure the area-averaged intensity of the PE and LW bands as a function of $\mucos = \khat \cdot \zhat$.
We then calculate radiation energy density by integrating the intensity with the mean density averaged horizontally at each $z$.
We adopt $\zpp=300\pc$.
This approach gives the mean radiation field as a function of $z$, which is uniform horizontally at a given $z$.
It is generally adequate for high-$z$ regions where the majority of gas is diffuse.
For the LyC band, we do not apply the condition (3), nor do we adopt the plane-parallel approximation at large $|z|$.

The background CR ionization rate is scaled relative to the solar neighborhood level based on the SFR. Specifically, we adopt $\xicrunatt = 2 \times 10^{-16} \,{\rm s}^{-1} \Sigma_{\rm SFR,40}'/\Sgas'$, where $\Sigma_{\rm SFR,40}'$ and $\Sgas'$ are the SFR surface density measured from stars formed in last 40 Myr and instantaneous gas surface density relative to solar neighborhood values $\Ssfr=2.5 \times 10^{-3} \sfrunit$ and $\Sgas=10\Surf$. The local CR ionization rate is then set to
\begin{equation}\label{e:cratt}
  \xicr =
  \begin{cases}
    \xicrunatt & \mbox{if } N_{\rm eff} \le N_{0} \,; \\
\xicrunatt \left(\frac{N_{\rm eff}}{N_{0}}\right)^{-1} &
\mbox{if } N_{\rm eff} > N_{0} \,,
\end{cases}
\end{equation}
where $N_{\rm eff}$ is the effective shielding column density and $N_{0} = 9.35\times 10^{20}\,\cm^{-2}$. To compute $N_{\rm eff}$ at each zone, we additionally follow the radiation energy density in the PE band without dust attenuation. We then use the ratio of attenuated to unattenuated PE radiation energy density to obtain
$N_{\rm eff} = 10^{21} \cm^{-2} Z_d'^{-1}\ln (\mathcal{E}_{\rm PE,unatt}/\mathcal{E}_{\rm PE})$.

We note that CR transport in the ISM is uncertain and our prescription for CR attenuation in setting $\xicr$ should be considered provisional; the term $1/\Sgas'$ in setting $\xicrunatt$ represents attenuation under average conditions and is motivated by \citet{2003ApJ...587..278W}. The additional attenuation at high column in Equation~\eqref{e:cratt} is motivated by observations of column-density dependent CR ionization rate \citep{2017ApJ...845..163N}.

\subsection{Photochemistry, Cooling, and Heating}\label{sec:pchem}

In the MHD integrator, we transport molecular and ionized hydrogen using passive scalars ($\xHH$ and $\xHII$, respectively) without source terms as implemented in {\tt Athena}.
We obtain the atomic hydrogen abundance from the closure relation $\xHI = 1 - 2\xHH - \xHII$.
We then update the temperature and abundances in an operator split manner by solving two ordinary differential equations (ODEs):
\begin{equation}\label{eq:ODE1}
  \frac{de}{dt} = K \dfrac{d T_{\mu}}{dt} = \mathcal{G} - \mathcal{L}\,,
\end{equation}
\begin{equation}\label{eq:ODE2}
  \frac{d x_s}{dt} = \mathcal{C}_s \,, \;\;\;\; \text{(s: ${\rm H}_2$ and ${\rm H}^+$)} \,,
\end{equation}
where $e=P/(\gamma - 1)$ is the internal energy density, $T_{\mu} \equiv T/\mu$ for $\mu = \muH/(1.1 + \xe - \xHH)$ the mean mass per particle, and $K = \nH \muH k_B/(\gamma-1)$ is taken as a constant. Given the generally short cooling/heating and chemical time scales,
in integrating these ODEs we take substeps (relative to the MHD time step) with the time step size determined by the minimum of MHD time step and 10\% of the cooling, heating, and chemical time scales.

At each substep, we solve the two ODEs sequentially. \autoref{eq:ODE1} is solved using the first-order backward difference formula with a Taylor expansion of the source terms, $\mathcal{G}-\mathcal{L}$, with respect to temperature which depends on the previous step's abundances and other information (see parameter dependence in \autoref{eq:Gamma_def} and \autoref{eq:Lambda_def} as well as Section 4 of \citet{KGKO}). We then evaluate $\mathcal{C}_s$ (see Section 3.1 of \citet{KGKO}) and solve \autoref{eq:ODE2} by treating it as a system of linear ODEs and use the backward Euler method. After the time-dependent update of hydrogen species, we compute the abundances of O$^+$, C$^+$, CO, C, and O under the steady-state assumption (see Sections 3.2 and 3.3 of \citet{KGKO}; see also \citealt{2017ApJ...843...38G}). We finally calculate the electron abundance $x_e$ from H$^+$ with contributions from C$^+$, O$^+$, and molecules at $T<2\times10^4\Kel$; or from He and metals assuming CIE at $T>2\times10^4\Kel$.

We refer the readers \citet{KGKO} for complete information on the processes we include and rates we adopt. Here, we list the formation and destruction processes that are explicitly considered, as well as the cooling and heating processes.
\begin{itemize}
    \item {\bf Molecular hydrogen:}
    formation by grain catalysis; destruction by CR ionization, photodissociation, photoionization, and collisional dissociation.
    \item {\bf Ionized hydrogen:} formation by
    photoionization, CR ionization, and collisional ionization of atomic hydrogen;
    destruction by radiative recombination and grain-assisted recombination.
    \item {\bf Ionized Carbon:} formation by photoionization, CR-induced photionization, and CR ionization of atomic carbon; destruction by grain-assisted recombination, radiative+dielectronic recombination, and the CH${}_2^{+}$ formation reaction.
    \item {\bf Heating:}
    photoelectric effect on small grains by FUV photons; CR ionization of $\Hn$ and $\HH$; photoionization of $\Hn$ and $\HH$; formation, photodissociation, and UV pumping of $\HH$.
    \item {\bf Cooling}:
    \begin{itemize}
        \item $\Lambda_{\rm hyd}$:
        collisionally-excited Ly$\alpha$ resonance line from $\Hn$; collisional ionization of $\Hn$; collisional dissociation of $\HH$; ro-vibrational lines from $\HH$; bremsstrahlung and radiative/grain assisted recombination of free electrons with $\Hp$.
        \item $\Lambda_{\rm others} ( T<2\times10^4\Kel)$:
        fine structure lines from C$^+$, C, and O; rotational lines of CO; combined nebular lines in ionized gas ($\Lambda_{\rm neb}$); grain-assisted recombination.
        \item $\Lambda_{\rm CIE} ( T>3.5\times10^4\Kel)$:
        ion-by-ion CIE cooling table for He and metals from \citet{2012ApJS..199...20G}. Metal cooling is scaled linearly with $Z_g'$.
    \end{itemize}
    We smoothly transition from $\Lambda_{\rm others}$ to $\Lambda_{\rm CIE}$ at $2\times10^4\Kel < T<3.5\times10^4\Kel$ using a sigmoid function, while $\Lambda_{\rm hyd}$ is applied at all temperatures using time-dependent hydrogen abundances.
\end{itemize}

\REV{}{We note that for the dust-associated process, we adopt an empirically constrained dust population model of \citet{Weingartner01a}, which consists of a separate population of carbonaceous and silicate grains as well as very small grains, including polycyclic aromatic hydrocarbon molecules.
Our standard choice is their model with grain size distribution A, $R_V=3.1$, and $b_C=4.0\times10^{-5}$.
}

\subsection{Models}\label{sec:model}

\begin{deluxetable}{lCCCCCC}
\tablecaption{Input Physical Parameters\label{tbl:model}}
\tablehead{
\colhead{Model} &
\dcolhead{\Sigma_{\rm gas,0}} &
\dcolhead{\Sstar} &
\dcolhead{\rho_{\rm dm}} &
\dcolhead{\Omega} &
\dcolhead{z_*} &
\dcolhead{R_0}
}
\colnumbers
\startdata
{\tt R8}    &  12 &   42 & 6.4\cdot 10^{-3}& 28 & 245  &  8  \\
{\tt LGR4}  &  50 &   50 & 5.0\cdot 10^{-3}& 30 & 500  &  4
\enddata
\tablecomments{
Column (2): initial gas surface density in $M_\odot\pc^{-2}$.
Column (3): stellar surface density in $M_\odot\pc^{-2}$.
Column (4): dark matter volume density at the midplane in $M_\odot\pc^{-3}$.
Column (5): angular velocity of galactic rotation at the domain center in ${\rm km\, s^{-1}\, kpc^{-1}}$.
Column (6): scale height of the stellar disk in pc.
Column (7): nominal galactocentric radius in kpc.
}
\end{deluxetable}

We consider two galactic conditions {\tt R8} and {\tt LGR4}, as described in \autoref{tbl:model}, which are analogous to the models of the same names in the TIGRESS-classic suite \citep[][the ``LG'' stands for the model with lower external gravity at a given gas surface density]{2020ApJ...900...61K}.
The {\tt R8} model represents conditions similar to the solar neighborhood \citep[e.g.,][]{2015ApJ...814...13M}. In terms of gas and stellar surface densities, the conditions in models {\tt R8} and {\tt LGR4} roughly correspond to the area-weighted and molecular-mass-weighted averages of conditions in  nearby star-forming galaxies  surveyed as part of PHANGS \citep{2022AJ....164...43S}.

For both simulations, the domain is a tall box, with dimensions $(1024\pc)^2\times 6144 \pc$ for {\tt R8}, and  $(512\pc)^2 \times 3072 \pc$ for {\tt LGR4}. We use $\dxymax = 2048\pc$ and $\epp=0$ for {\tt R8} and $\dxymax=1024\pc$ and $\epp=10^{-8}$ for {\tt LGR4}. With these choices of numerical parameters  for ray-tracing, we found good convergence for the EUV radiation field everywhere, the FUV field near the midplane, and overall simulation outcomes (see \autoref{sec:A-rt-conv}). We note that the selection of the optimal values of $\dxymax$ and $\epp$ depends on the system's conditions, especially on $\Zd$  (which determines dust absorption).

We run each simulation with two different resolutions: $8\pc$ and $4\pc$ for {\tt R8}; and $4\pc$ and $2\pc$ for {\tt LGR4}. The initial gas distribution follows double Gaussian profiles \citep[see][]{2017ApJ...846..133K} representing warm and hot components with the Gaussian scale height corresponding to initial velocity dispersions of 10 and 100~km/s for {\tt R8} and 15 and 150~km/s for {\tt LGR4}. To reduce initial transients, we add additional velocity perturbations with amplitude of 10 and 15 km/s for {\tt R8} and {\tt LGR4}, respectively, along with randomly distributed initial star clusters that provide non-zero initial heating and SNe. The initial magnetic field is set to be azimuthal $\Bvec=B_0\yhat$ with uniform plasma beta $\beta_0\equiv \Pth/(B_0^2/8 \pi)=1$, which is close to the expected saturation value of the magnetic field \citep[e.g.][]{2015ApJ...815...67K,2022ApJ...936..137O}. After one or two cycles of star formation and feedback, the system reaches a quasi-steady state  with minimal impact from initial transients and the choice of density profiles and the level of initial turbulence.

\section{TIGRESS-NCR Simulations}\label{sec:simulations}

We first provide an overview of the {\tt R8} and {\tt LGR4} simulations. Here, we focus on global ISM properties and energetics as well as the distribution of the multiphase (both thermally and chemically) ISM near the galactic midplane.

\begin{deluxetable*}{lCCCCCCC}
\tablecaption{Global Properties at Saturation\label{tbl:satprop}}
\tablehead{
\colhead{Model} &
\dcolhead{\Sgas} &
\dcolhead{{\Ssfr}_{,10}} &
\dcolhead{H} &
\dcolhead{\sigma_{{\rm eff}}} &
\dcolhead{\sigma_{z, {\rm turb}}} &
\dcolhead{t_{\rm ver}} &
\dcolhead{t_{\rm dep}} \\
\colhead{} &
\colhead{$(M_\odot\pc^{-2})$} &
\colhead{$(10^{-3} \sfrunit)$} &
\colhead{(pc)} &
\colhead{(km/s)} &
\colhead{(km/s)} &
\colhead{(Myr)} &
\colhead{(Gyr)}
}
\colnumbers
\startdata
{\tt R8-4pc}     & 10.6^{+0.2}_{-0.2} &  2.8^{+1.5}_{-1.0} & 199.3^{+36.2}_{-44.3} & 12.0^{+1.0}_{-0.5} &  7.7^{+1.0}_{-0.7} & 23.3^{+8.0}_{-3.8} &  3.6^{+1.0}_{-1.0} \\
{\tt R8-8pc}     & 10.3^{+0.4}_{-0.2} &  2.8^{+3.5}_{-2.0} & 233.5^{+147.1}_{-57.9} & 13.5^{+2.9}_{-1.3} &  9.4^{+3.3}_{-2.1} & 24.3^{+16.7}_{-5.2} &  3.2^{+2.5}_{-1.4} \\
{\tt LGR4-2pc}   & 37.9^{+1.3}_{-0.9} & 34.8^{+10.4}_{-10.7} & 164.5^{+31.4}_{-47.1} & 13.4^{+0.7}_{-0.7} &  8.3^{+0.9}_{-0.8} & 17.8^{+6.7}_{-3.9} &  1.2^{+0.2}_{-0.2} \\
{\tt LGR4-4pc}   & 36.2^{+1.4}_{-0.6} & 29.0^{+20.3}_{-8.5} & 176.2^{+64.0}_{-66.2} & 14.6^{+1.2}_{-1.1} &  9.9^{+1.5}_{-1.2} & 15.4^{+9.6}_{-4.0} &  1.0^{+0.4}_{-0.1} \\
\enddata
\tablecomments{
Each column provides median values as well as the 16$^{\rm th}$ to 84$^{\rm th}$ percentile range, over $t=250-450\Myr$ for {\tt R8} and $t=250-350\Myr$ for {\tt LGR4}.
Column (2): gas surface density.
Column (3): SFR surface density.
Column (4): mass-weighted gas scale height.
Column (5): effective vertical velocity dispersion.
Column (6): turbulent component of vertical velocity dispersion.
Column (7): vertical dynamical time scale.
Column (8): gas depletion time.
Note that Columns (4)-(7) are calculated for the warm and cold gas with temperature $T<3.5\times10^4\Kel$.
See text for definitions.}
\end{deluxetable*}

\begin{figure*}
    \centering
    \includegraphics[width=\linewidth]{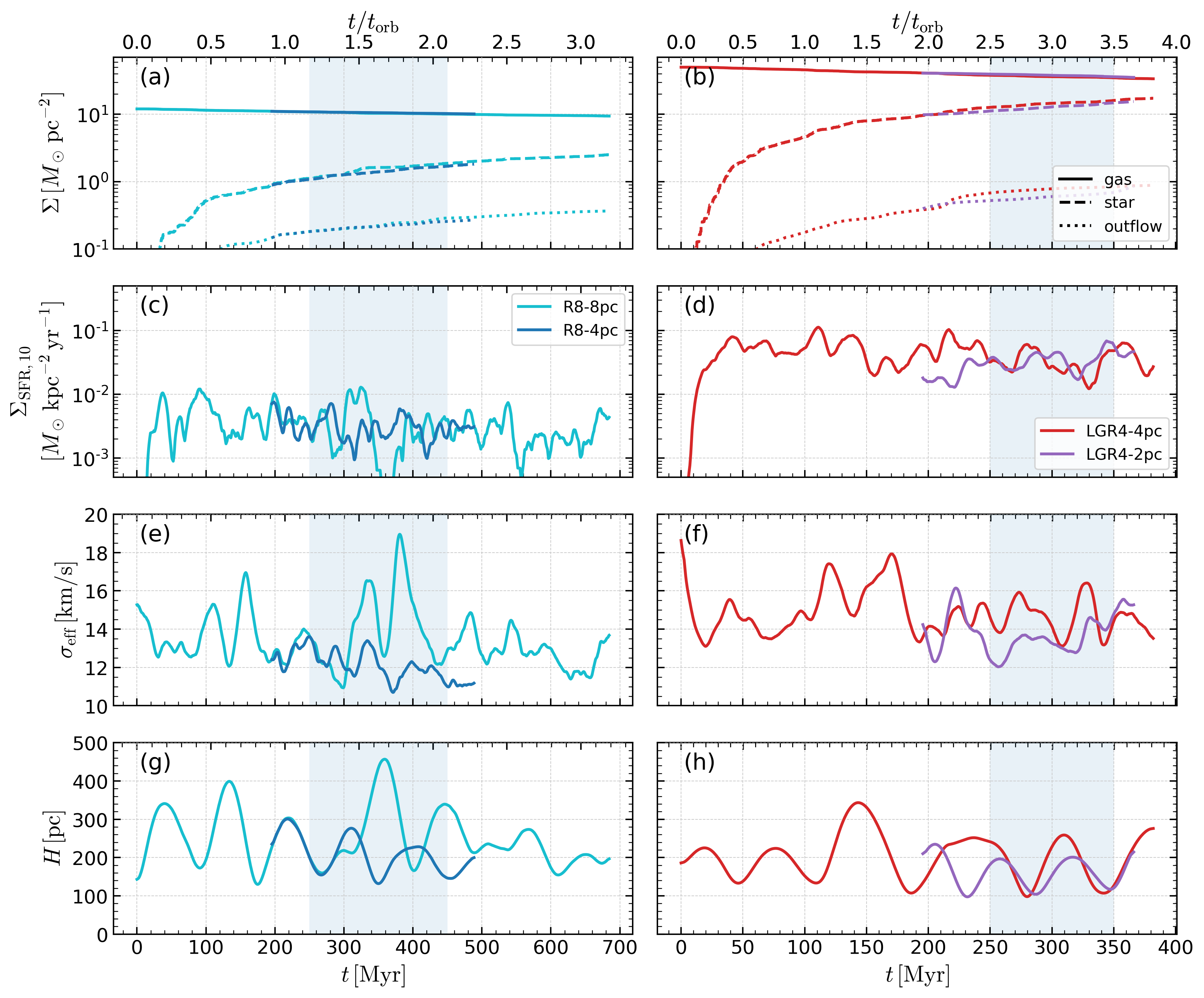}
    \caption{Time evolution of global properties in model {\tt R8} (left) and {\tt LGR4} (right). From top to bottom, we plot (a) \& (b) surface densities of gas, newly-formed stars, and outflows, (c) \& (d) SFR surface density (over last 10Myr), (e) \& (f) effective vertical velocity dispersion, and (g) \& (h) gas scale height. Results from models with different resolutions are presented, as noted in the keys. We apply 5~Myr rolling averages to reduce high-frequency fluctuations in order to ease comparison between different resolution models. The shaded area represents the time interval over which the saturated properties are calculated.}
    \label{fig:tevol_global}
\end{figure*}

\begin{figure*}
    \centering
    \includegraphics[width=\textwidth]{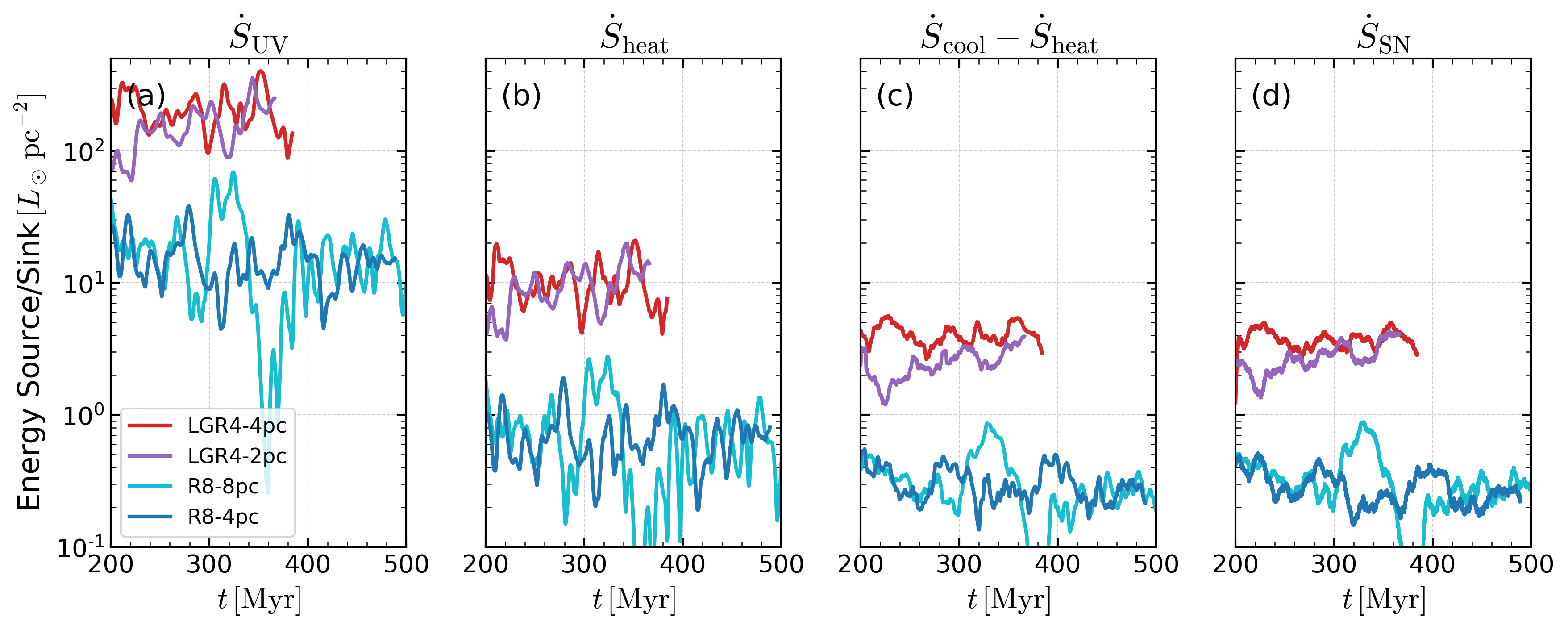}
    \caption{Time evolution of energy source and sink rates per area in the simulation. From left to right, we show (a) the UV radiation energy injection rate, (b) the radiative heating rate, (c) the net cooling rate, and (d) the SN energy injection rate. About $5\%$ of UV radiation energy goes into heating the warm and cold gas. The total radiative cooling always exceeds radiative heating because the cooling offsets heating provided by SNe. Only 2--3\% of the injected SN energy \REV{leaves}{is advected out} of our simulation domain.}
    \label{fig:tevol_ss}
\end{figure*}

\begin{figure*}
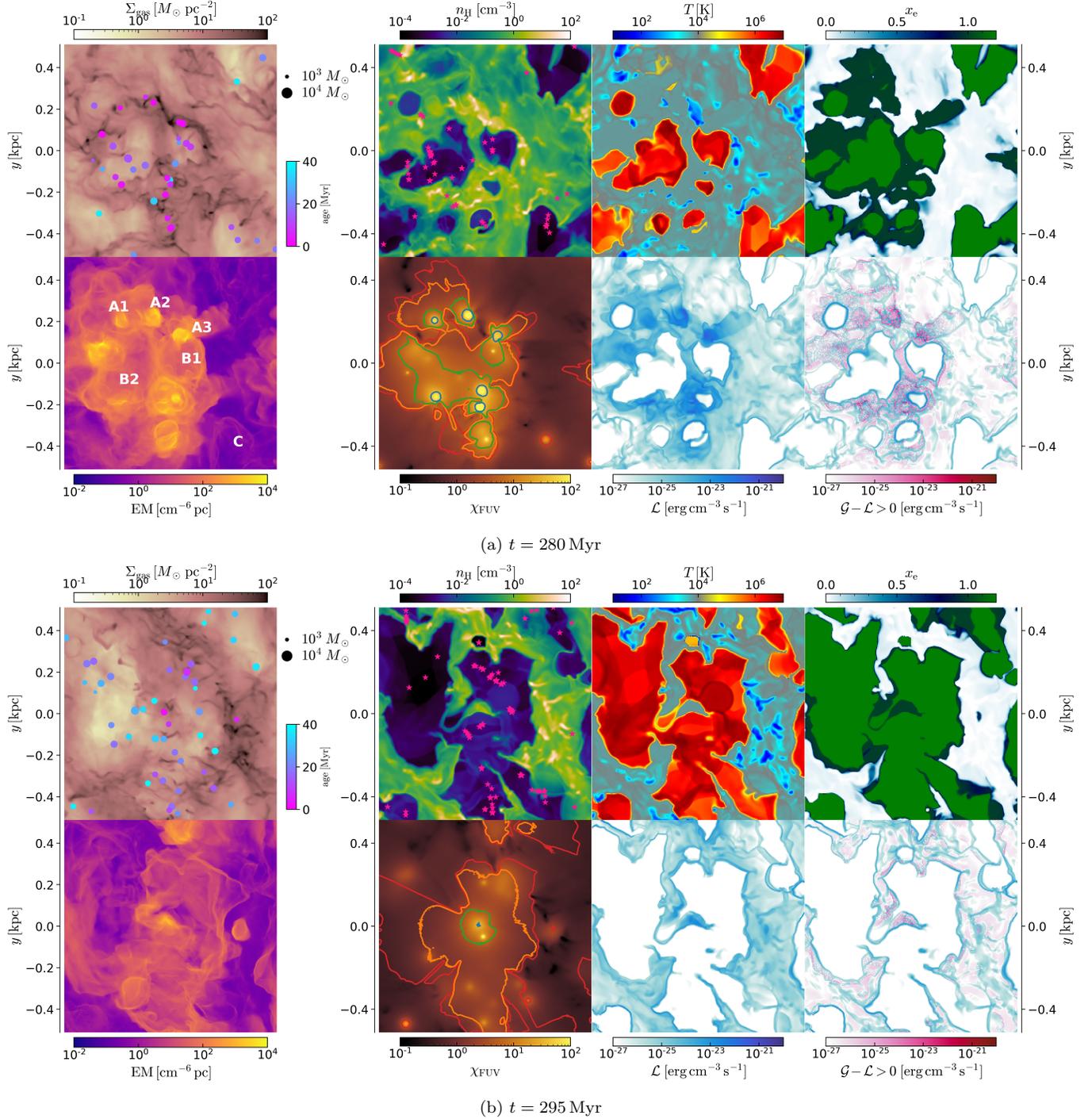

    \centering
    \fig{{R8_4pc_NCR.full.xy2048.eps0.np768.has.0285.snapshot}.png}{\textwidth}{(a) $t=280\Myr$}
    \fig{{R8_4pc_NCR.full.xy2048.eps0.np768.has.0300.snapshot}.png}{\textwidth}{(b) $t=295\Myr$}
    \caption{Visualization of the simulated ISM from model {\tt R8-4pc} at (a) $t=280\Myr$ -- top set of panels; and (b) $295\Myr$ -- bottom set of panels. {\bf Left:} Gas surface density \REV{}{(top)} and emission measure \REV{}{(bottom)} in the $x$-$y$ plane (corresponding to integrals of $\rho$ and $n_e^2$ along the $z$-axis, \REV{}{respectively}). The letters in the emission measure map indicate regions of ionized bubbles (warm ionized bubble -- A1, A2, A3; young superbubble -- B1, B2; old superbubble -- C).
    {\bf Right:} Slices through the midplane, $z=0$. From left to right, the top row shows hydrogen number density $\nH$, temperature $T$, and electron fraction $x_e$;
    the bottom row shows normalized FUV radiation energy density $\chiFUV$, cooling rate $\mathcal{L}$, and net heating rate $\mathcal{G}-\mathcal{L}$. $\chi_{\rm FUV}\equiv J_{\rm FUV}/J_{\rm FUV}^{\rm D78}$, where $J_{\rm FUV}$ is the FUV mean intensity and $J_{\rm FUV}^{\rm D78} = 2.1\times10^{-4}\junits$ \citep{1978ApJS...36..595D}. Note that for $\mathcal{G}-\mathcal{L}$, the pink colormap is used only for positive values (net heating), while the blue colormap from  $\mathcal{L}$ is used for negative values (net cooling). Contours of EUV radiation energy density are overlaid in the  $\chiFUV$ panels for $\log [\mathcal{E}_{\rm LyC}/(\erg\pcc)] = -15$ (red), $-14$ (orange), $-13$ (green), and $-12$ (blue).
    In the $\Sgas$ panel, young star clusters with age $<40$~Myr and $|z|<50\pc$ are shown as circles. The size of circles is proportional to the cluster mass, but in practice its range is narrow ($10^3-5\times10^3\Msun$).
    Clusters with age $<20$~Myr (magenta-ish colors; see colormap in the bottom-right of the $\Sgas$ panel) are FUV sources, while very young clusters (age $<5$~Myr) are the only significant EUV sources (enclosed by green/blue contours in the $\chi_{\rm FUV}$ panel).
    Locations where SNe exploded within the past 10~Myr, and within $|z|<15\pc$ of the slice shown,
    are marked as star symbols in the $\nH$ panel.}
    \label{fig:R8_snapshot}
\end{figure*}

\subsection{Global ISM Properties}\label{sec:ism_prop}

Figure~\ref{fig:tevol_global} shows the time evolution of key global quantities including
(a) the gas surface density $\Sgas\equiv M_{\rm gas}/(L_xL_y)$ along with surface densities of
newly formed stars $\Sstarnew\equiv M_{\rm *, new}/(L_xL_y)$ and mass loss via outflows  from
the computational domain $\Sout\equiv M_{\rm out}/(L_xL_y)$;
(b) the SFR surface density over the last $\Delta t = 10 \Myr$
\begin{equation}\label{eq:sfr10}
{\Ssfr}_{, 10}\equiv \frac{\sum_i M_{*,i}(t_{\rm age}<\Delta t)}{L_xL_y\Delta t},
\end{equation}
where $M_{*,i}(t_{\rm age}<\Delta t)$ is the total mass of star particles with age younger than $\Delta t$;
(c) the effective vertical velocity dispersion
\begin{equation}\label{eq:szeff}
    \sigma_{{\rm eff}} \equiv \sbrackets{ \frac{\int\rbrackets{\rho v_z^2 + P + B^2/(8\pi) - B_z^2/(4\pi)}dV}{\int\rho dV}}^{1/2};
\end{equation}
and (d) the mass-weighted scale height of the gas
\begin{equation}\label{eq:H}
    H\equiv\rbrackets{\frac{\int \rho z^2 dV}{\int \rho dV}}^{1/2}.
\end{equation}
Here, the values of $\sigma_{\rm eff}$ and $H$ are calculated only for the warm and cold gas with temperature $T<3.5\times10^4\Kel$.
\REV{because these quantities for the hot gas are not well defined in local simulations.}{}
We discuss phase separated velocity dispersions in \autoref{sec:phase}.
We note that the magnetic term in  $\sigma_{{\rm eff}}$ is not simply the magnetic pressure, but instead represents the vertical component of Maxwell stress including both magnetic pressure and tension. We measure the mass-weighted vertical velocity dispersion of the turbulence for the warm and cold gas, i.e. $\sigma_{z, {\rm turb}}\equiv (\int \rho v_z^2dV/\int\rho dV)^{1/2}$, and report this separately in \autoref{tbl:satprop}.

The simulations begin with initial rough hydrostatic equilibrium with $H\sim 150-200\pc$. The idealized initial setups soon lead to a burst of star formation and associated feedback as the disk loses its initial vertical support from both thermal and turbulent pressure. During the first $\sim 100\Myr$ evolution, both models experience at least two complete star formation and feedback cycles (rise and fall of SFRs), whose timescale is proportional to the vertical crossing timescale of the disk \citep{2020ApJ...900...61K}.
To reduce the computational time needed for high-resolution models, we refine and restart low-resolution simulations ({\tt R8-8pc} and {\tt LGR4-4pc}) after the initial transient has passed ($t=200\Myr$).
The mesh-refined, restarted models are run for longer than one orbit time (or 3-4 star formation and feedback cycles) to obtain a fair statistical sample of states at higher resolution. In \autoref{tbl:satprop}, we summarize mean and standard deviation of quantities of interest over $t=250-450\Myr$ and $t=250-350\Myr$ for {\tt R8} and {\tt LGR4}, respectively, at different resolutions.
Our results verify the overall convergence of the global properties with respect to numerical resolution.\footnote{\REV{}{We note that the star-forming ISM is inherently stochastic, showing large variation among different realizations. For example, in {\tt R8} the lower resolution model ({\tt R8-8pc}) shows a big burst at 300~Myr, which is absent in {\tt R8-4pc}. For a stricter convergence test, a statistical comparison between many realizations would be required, which is too computationally costly to be practical at present. For now, our statement regarding resolution convergence is based on the median is well within the time variation of quantities.}}

As shown in the top row of \autoref{fig:tevol_global}, the gas surface density (solid) decreases gradually due to star formation (dashed) and outflows (dotted).
The global properties shown in the bottom three rows of \autoref{fig:tevol_global} reach a quasi-steady state, with substantial temporal fluctuations
($\sim 0.2-0.3$ dex), and show quasi-periodic fluctuations.
The characteristic period is the vertical oscillation time determined by the \emph{total} (gas+star+dark matter) midplane density $\sim (4\pi G \rho_{\rm tot})^{-1/2}$, which is similar to the vertical crossing time \citep[see][]{2020ApJ...900...61K}. In \autoref{tbl:satprop}, we list the vertical crossing time $t_{\rm ver}\equiv H/\sigma_{\rm z, turb}$ and gas depletion time $t_{\rm dep}\equiv \Sgas/\Ssfr$. Quasi-periodic fluctuating behavior in $\Ssfr$ and $\sigma_{{\rm eff}}$ shows higher frequency fluctuations than $H$.
Occasionally, when there is a big burst, systematic offsets among three quantities stand out; a peak of SFR is followed by a peak of velocity dispersion and then scale height (e.g., see peaks near 100~Myr and 300~Myr in {\tt R8-8pc}).

\subsection{Global Energetics}\label{sec:energetics}

\autoref{fig:tevol_ss}(a) shows
\REV{the time evolution of}{}
the total energy injection rate per unit area by UV radiation\footnote{Here, we use the $\dot{S}$ notation for any energy gain and loss rates per unit area. In previous publications \citep[e.g.,][]{2020ApJ...900...61K,2022ApJ...936..137O}, we used $\Sigma_{\rm FUV}$ for the surface density of FUV luminosity. With the current notation, $\dot{S}_{\rm FUV}\equiv\Sigma_{\rm FUV}$.}
$\dot{S}_{\rm UV} \equiv L_{\rm UV,tot}/(L_x L_y)$  \REV{}{as a function of time}, where $L_{\rm UV,tot}$ is the total UV (PE+LW+LyC) luminosity of star particles with $t_{\rm age} < 20\Myr$.
This energy injection rate is
\REV{solely}{}
determined by the adopted SPS model (STARBURST99 in our case) and recent star formation history.

UV radiation propagates through the ISM and is absorbed by gas and dust, photoionizing some regions where EUV penetrates, and heating up the gas via the photoelectric effect in other regions where FUV penetrates. In addition, CR ionization is an important heating source in regions that are shielded to both EUV and FUV.
The total \emph{radiative} (including CR) heating rate per unit area
$\dot{S}_{\rm heat} \equiv\int \mathcal{G} dV/(L_xL_y)$
shown in \autoref{fig:tevol_ss}(b) is the sum of hydrogen photoionization heating by LyC radiation ($\dot{S}_{\rm heat,PI}/ \dot{S}_{\rm heat}\sim 75\%$), photoelectric heating from FUV (PE+LW) radiation on small dust grains ($\dot{S}_{\rm heat,PE}/ \dot{S}_{\rm heat} \sim 20\%$), and CR ionization heating ($\dot{S}_{\rm heat,CR}/ \dot{S}_{\rm heat} \sim 1-2\%)$, plus a tiny contribution from $\HH$ heating ($<0.1\%$). The global heating efficiency, defined as the ratio of the total heating rate to the UV radiation injection rate, is
$\dot{S}_{\rm heat,PI+PE}/\dot{S}_{\rm UV}\sim 5-6\%$, with individual efficiencies
$\dot{S}_{\rm heat,PE} / \dot{S}_{\rm FUV}\sim 2\%$  and
$\dot{S}_{\rm heat,PI} / \dot{S}_{\rm EUV}\sim 15\%$.

The radiative heating within the simulation domain is balanced by radiative cooling. \autoref{fig:tevol_ss}(c) shows the difference between cooling and heating per unit area within the simulation volume.
The difference is positive, indicating net cooling. The excess in radiative cooling is offset by energy input from SN feedback ($\dot{S}_{\rm SN}$; \autoref{fig:tevol_ss}(d)).
$\dot{S}_{\rm SN}$ is about two orders of magnitude smaller than the UV radiation injected ($\dot{S}_{\rm UV}$; \autoref{fig:tevol_ss}(a)), and a factor $\sim3$ lower than the radiative heating rate $\dot{S}_{\rm heat}$.
A small fraction of energy also leaves the computational domain through outflows; the majority of outflowing energy is in the hot gas and therefore originally deposited by SNe, and the kinetic energy of outflowing gas is also powered by SNe. This outflowing energy escaping the domain ($\sim2-3\%$ of $\dot{S}_{\rm SN}$) accounts for the small excess of SN injection energy over the net cooling within the domain.

\subsection{ISM Cartography}\label{sec:cartography}

The instantaneous spatial distribution of the ISM mass and energy densities is highly structured and complex. To provide a visual impression of the ISM structure in our simulations, we display maps of various quantities from {\tt R8-4pc} at two epochs, $t=280$ and 295~Myr in \autoref{fig:R8_snapshot}(a) and (b). These times respectively correspond to a local peak and trough of $\Ssfr$ (see \autoref{fig:tevol_global}(c)).
We note that qualitative features presented here for {\tt R8} are also seen in {\tt LGR4}.

We first focus on the epoch shown in \autoref{fig:R8_snapshot}(a), shortly after a burst of star formation \REV{}{has} occurred, during which many new star clusters were formed. Very young clusters ($t_{\rm age} < 5\Myr$) act as strong UV radiation sources. These clusters are spatially correlated with each other and with the dense clouds where they were born.
There are two distinct types of bubble structures: hot SN-driven bubbles (characterized by low density, diffuse EM, and high temperature) and warm ionized bubbles (characterized by high EM). The electron fraction of the two types of bubbles are different. The hot ionized gas has higher $\xe\approx 1.2$ (bright green), due to free electrons from collisionally ionized H, He, and metals.
In contrast, $\xe\approx 1$ (dark green) in the warm ionized gas as electrons are mostly from photoionized H (and a tiny contribution from $\mathrm{C^+}$, $\mathrm{O^+}$, and molecular ions). In the upper region of the domain ($y\sim 0.2\kpc$), examples of warm ionized bubbles, corresponding to high-EM sites, are marked as A1, A2, and A3. In the middle region ($y\sim 0\kpc$), two superbubbles that are formed relatively recently and show moderate EM are marked as B1 and B2. An example of an old, low EM superbubble is marked as C.

It is evident that recently born star cluster complexes are responsible for photoionizing bubbles A1, A2, and A3. Ionizing radiation from clusters near A1 and A2 is fairly well blocked toward bubbles B1 and B2, while extended radiation from these sources could ionize a substantial area toward the top of the domain. Large areas within the domain remain neutral ($x_e\lesssim 10\%$; white-blue in the $\xe$ panel) as EUV is effectively truncated due to the large cross section of the neutral hydrogen. FUV is only significantly absorbed by dense clouds, making the radiation energy density low in their interiors and also casting shadows behind them. Still, it is evident that most of the neutral gas is irradiated with FUV ($\chi_{\rm FUV}\gtrsim 0.5$), which is a major heating source. Bubble C is an old, hot superbubble, and star clusters are old and no longer contributing significant EUV. As a result, the hot gas is bounded by the neutral gas. In contrast, the intermediate age clusters inside hot bubbles B1 and B2, together with other nearby young clusters, create a photoionized region between the hot and neutral gas. The strongest cooling in the slice, as shown in the $\cal{L}$ panel, occurs in the photoionized regions near Bubbles A and B; the main coolants are nebular lines of metal ions. However, the heating produced by ionizing radiation offsets or even exceeds the cooling in this region, leading to net heating (see the $\cal{G}-\cal{L}$ panel). The net cooling rate is highest at hot bubble boundaries (CIE cooling at $T\sim10^5\Kel$).
These interfaces where hot gas mixes with denser gas to become strongly radiative are important in bubble energetics.

As young star clusters shown in \autoref{fig:R8_snapshot}(a) age, they begin to produce SNe, resulting in superbubbles, which merge into a very large hot bubble in the center of \autoref{fig:R8_snapshot}(b). This is carved by several clustered SNe (positions are shown as star symbols in the $\nH$ panel). At this epoch, there is only one significant ionizing source at the center of the midplane slice, and the area filled with the warm ionized gas (dark green in the $\xe$ panel) is greatly reduced. There are a few out-of-midplane sources (not shown in the $\Sgas$ panel) responsible for large EM bubbles at $(x,y)\sim(0, 0.5)\kpc$ and $(x,y)\sim (-0.2,-0.3)\kpc$. It is also noticeable that old clusters are now spread across the simulation domain; clustering of clusters is reduced.

There is temporal evolution in the ISM phase structure over the interval shown between the two snapshots. The midplane volume filling factor of the warm ionized medium achieves its local maximum, $\sim 30\%$, at $t=280\Myr$ (\autoref{fig:R8_snapshot}(a)), decreasing to $\sim 10\%$ within another 5~Myr. In contrast, the midplane hot gas filling factor increases gradually from $20\%$ at $t=280\Myr$ to $50\%$  at $t=295\Myr$. The filling factor of the warm neutral medium changes from $40\%$ to $20\%$ over the same $15\Myr$ interval.

\begin{deluxetable*}{lCCc}
\tablecaption{Phase Definition \label{tbl:phase}}
\tablehead{
\colhead{Name} &
\dcolhead{{\rm Temperature}} &
\dcolhead{{\rm Abundance}} &
\colhead{{\rm Shorthand}}
}
\startdata
Cold Molecular Medium\tablenotemark{a} & T<6\cdot10^3\Kel & \xHH>0.25 & \CMM \\
Cold Neutral Medium\tablenotemark{b} & T<500\Kel & \xHI>0.5 & \CNM \\
Unstable Neutral Medium & 500\Kel<T<6\cdot10^3\Kel & \xHI>0.5 & \UNM \\
Unstable Ionized Medium\tablenotemark{c} & T<6\cdot10^3\Kel & \xHII>0.5 & \UIM \\
Warm Neutral Medium & 6\cdot10^3\Kel<T<3.5\cdot10^4\Kel & \xHI>0.5 & \WNM \\
Warm Photo Ionized Medium & 6\cdot10^3\Kel<T<1.5\cdot10^4\Kel & \xHII>0.5 & \WPIM \\
Warm Collisionally Ionized Medium & 1.5\cdot10^4\Kel<T<3.5\cdot10^4\Kel & \xHII>0.5 & \WCIM \\
Warm-Hot Ionized Medium & 3.5\cdot10^4\Kel<T<5\cdot10^5\Kel & \cdots & \WHIM \\
Hot Ionized Medium & 5\cdot10^5\Kel<T & \cdots & \HIM \\
\enddata
\tablenotetext{a}{This includes unstable temperature range but is dominated by cold.}
\tablenotetext{b}{In principle, `neutral' includes both `atomic' and 'molecular'. But historically, the cold neutral medium has been used to denote cold atomic medium. Here, we follow the convention.}
\tablenotetext{c}{This includes cold temperature range but is dominated by unstable.}
\end{deluxetable*}

\subsection{Phase Definition, Filling Factor, and Velocity Dispersion}\label{sec:phase}

\begin{figure*}
    \centering
    \includegraphics[width=\textwidth]{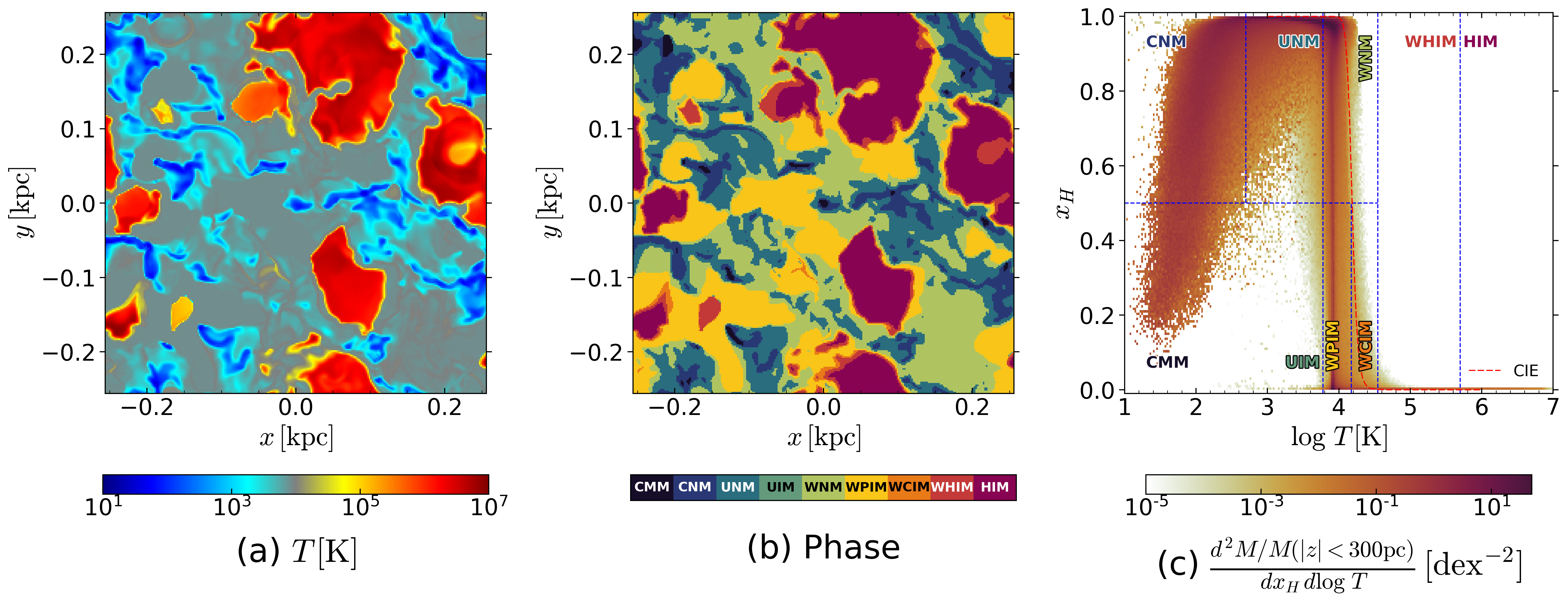}
    \caption{Example showing the gas phase distribution from {\tt LGR4-2pc} at $t=230\Myr$.
    (a) Midplane slice of temperature.
    (b) Regions assigned to mutually exclusive defined phases as shown in the key. (c) Mass-weighted joint PDF of $\log\,T$ and $\xHI$ for gas within $|z|<300\pc$, with dividing lines for different gas phase definitions (see \autoref{tbl:phase}). The red dashed curve in panel (c) denotes the relation between $\xHI$ and $T$ based on CIE of hydrogen.
    \label{fig:phase}}
\end{figure*}

\begin{deluxetable*}{lCCCCCCC}
\tablecaption{Mass Fractions and Volume Filling Factors with $|z|<300\pc$ \label{tbl:filling}}
\tablehead{
\colhead{Model} &
\colhead{\Cold} &
\colhead{\UNM} &
\colhead{\WNM} &
\colhead{\WPIM} &
\colhead{\WCIM} &
\colhead{\WHIM} &
\colhead{\HIM}
}
\startdata
\multicolumn{8}{c}{Mass fractions} \\
{\tt R8-4pc}     & 27.4^{+4.8}_{-10.3} & 28.8^{+3.4}_{-3.1} & 34.9^{+10.6}_{-5.9} & 7.5^{+5.6}_{-3.1} & 0.2^{+0.2}_{-0.1} & 0.1^{+0.1}_{-0.0} & 0.05^{+0.05}_{-0.02} \\
{\tt R8-8pc}     & 22.0^{+11.6}_{-10.8} & 29.1^{+5.6}_{-8.7} & 36.0^{+15.0}_{-9.6} & 6.7^{+10.6}_{-5.4} & 0.3^{+0.3}_{-0.1} & 0.2^{+0.2}_{-0.1} & 0.07^{+0.09}_{-0.03} \\
{\tt LGR4-2pc}   & 37.3^{+3.2}_{-5.3} & 27.4^{+1.3}_{-0.9} & 27.2^{+3.1}_{-3.3} & 7.3^{+2.3}_{-1.7} & 0.2^{+0.1}_{-0.1} & 0.1^{+0.1}_{-0.0} & 0.07^{+0.03}_{-0.02} \\
{\tt LGR4-4pc}   & 31.6^{+3.8}_{-4.9} & 30.0^{+1.5}_{-1.3} & 29.9^{+4.5}_{-3.5} & 7.3^{+3.6}_{-2.1} & 0.3^{+0.1}_{-0.1} & 0.2^{+0.1}_{-0.1} & 0.10^{+0.03}_{-0.03} \\
\hline
\multicolumn{8}{c}{Volume filling factors} \\
{\tt R8-4pc}     & 1.6^{+0.5}_{-1.0} & 17.0^{+4.2}_{-6.0} & 48.2^{+6.9}_{-5.9} & 10.9^{+7.6}_{-4.4} & 1.4^{+0.3}_{-0.3} & 3.6^{+1.1}_{-0.9} & 12.5^{+15.5}_{-3.8} \\
{\tt R8-8pc}     & 1.5^{+1.9}_{-1.1} & 14.4^{+8.5}_{-9.3} & 44.5^{+11.8}_{-11.9} & 10.0^{+12.5}_{-7.5} & 1.7^{+0.7}_{-0.5} & 4.0^{+1.4}_{-1.4} & 17.5^{+20.7}_{-9.3} \\
{\tt LGR4-2pc}   & 2.5^{+0.7}_{-0.6} & 14.5^{+4.2}_{-2.0} & 51.3^{+10.4}_{-8.4} & 9.3^{+3.2}_{-2.1} & 1.1^{+0.2}_{-0.3} & 3.1^{+0.7}_{-1.0} & 15.4^{+6.6}_{-8.9} \\
{\tt LGR4-4pc}   & 2.1^{+1.0}_{-0.6} & 12.6^{+3.9}_{-1.4} & 50.9^{+6.1}_{-8.2} & 9.8^{+4.1}_{-3.3} & 1.8^{+0.3}_{-0.3} & 3.7^{+0.7}_{-0.7} & 18.3^{+4.9}_{-7.2} \\
\enddata
\tablecomments{
Each column shows the median and 16$^{\rm th}$ and 84$^{\rm th}$ percentile range over $t=250-450\Myr$ and $t=250-350\Myr$ for {\tt R8} and {\tt LGR4}, respectively.
}
\end{deluxetable*}

\begin{deluxetable*}{lCCCCcCCCC}
\tablecaption{Mass-weighted vertical velocity dispersion \label{tbl:veld}}
\tablehead{
\multirow{2}{*}{Model} &
\multicolumn{4}{c}{$\sigma_{\rm eff}$} & &
\multicolumn{4}{c}{$\sigma_{\rm z,turb}$} \\
\cline{2-5}
\cline{7-10}
 &
\colhead{\Cold} &
\colhead{\UNM{}+\WNM} &
\colhead{\twop} &
\colhead{\WIM} & &
\colhead{\Cold} &
\colhead{\UNM{}+\WNM} &
\colhead{\twop} &
\colhead{\WIM}
}
\startdata
{\tt R8-4pc}     & 5.4^{+1.2}_{-0.9} & 12.6^{+0.6}_{-0.7} & 11.2^{+0.6}_{-0.6} & 18.4^{+2.1}_{-1.9} & & 4.2^{+1.5}_{-0.6} & 7.6^{+1.1}_{-0.7} & 6.9^{+1.0}_{-0.7} & 13.0^{+2.5}_{-2.0} \\
{\tt R8-8pc}     & 5.8^{+1.8}_{-0.7} & 13.1^{+2.2}_{-1.1} & 12.2^{+2.2}_{-1.5} & 21.1^{+10.6}_{-3.4} & & 4.8^{+1.5}_{-1.1} & 8.6^{+2.1}_{-1.8} & 8.1^{+2.5}_{-1.8} & 16.2^{+12.3}_{-4.1} \\
{\tt LGR4-2pc}   & 6.1^{+1.1}_{-0.7} & 14.8^{+0.6}_{-1.2} & 12.9^{+0.7}_{-0.7} & 18.7^{+1.1}_{-0.8} & & 5.3^{+1.4}_{-0.7} & 8.4^{+1.1}_{-0.7} & 7.7^{+1.1}_{-0.7} & 12.9^{+1.6}_{-1.3} \\
{\tt LGR4-4pc}   & 7.7^{+3.3}_{-1.2} & 15.9^{+1.4}_{-1.4} & 13.8^{+1.6}_{-1.2} & 21.0^{+2.0}_{-1.4} & & 6.7^{+3.2}_{-1.8} & 10.1^{+1.8}_{-1.3} & 9.2^{+2.1}_{-1.1} & 14.8^{+2.5}_{-1.6} \\
\enddata
\tablecomments{
Each column shows the median and 16$^{\rm th}$ and 84$^{\rm th}$ percentile range over $t=250-450\Myr$ for {\tt R8} and $t=250-350\Myr$ for {\tt LGR4}.
}
\end{deluxetable*}

Traditionally, in the ISM literature, the gas is often divided into different phases based on temperature and hydrogen chemical state.
We choose a set of specific temperature and abundance cuts to define 9 phases as summarized in \autoref{tbl:phase}.  The distributions of these phases are depicted for a sample snapshot from {\tt LGR4-2pc} in \autoref{fig:phase}.
In our previous work (\citealt{2017ApJ...846..133K} and subsequent works based on TIGRESS-classic), we used temperature cuts only to define 5 ISM phases. Key additional information available in the current study is the time-dependent hydrogen abundance, allowing for subdivisions, e.g., ``warm'' into the warm neutral and warm ionized medium. The warm ionized medium is further divided into ``warm-photoionized'' and ``warm-collisionally-ionized'' medium with a temperature cut at $T=1.5\cdot10^4\Kel$, above which hydrogen can be collisionally ionized (see red dashed line in \autoref{fig:phase}(c)).
We assign every cell to one of the 9 phases exclusively.

A summary of the mass and volume fraction for each phase is shown in \autoref{tbl:filling}, for both {\tt R8} and {\tt LGR4}.
Here, we do not explicitly distinguish \CNM{} and \CMM{} and
we ignore \UIM{} given its negligible mass and volume fractions. Instead, we use \Cold{} for the combined cold medium (\CMM+\CNM). We note that, as shown in \autoref{fig:phase}(c), hydrogen species abundances vary continuously, and a significant amount of \emph{partially} molecular gas is present in \CNM{}. The total molecular gas mass is thus larger than the mass of \CMM{}.
We note that the \Cold{} mass fraction increases substantially with higher numerical resolution at the expense of \UNM{} and \WNM{}, but the fractions of all the other phases are reasonably converged.
\WNM{} fills the majority of the volume, with substantial contributions from \WPIM{} and \HIM{} as well as \UNM{}. The neutral gas (\Cold{}, \UNM{}, and \WNM{}) dominates the total mass budget. \WPIM{} contributes to both volume and mass at $\sim 10\%$ level, with an increasing contribution at high altitudes (e.g., \citealt{2020ApJ...897..143K}; see also N. Linzer et al. in prep).

Separating the warm and cold gas into \Cold{}, \UNM+\WNM, \twop, and \WIM{} components, \autoref{tbl:veld} shows the effective vertical velocity dispersion as defined by \autoref{eq:szeff} and the mass-weighted turbulent velocity dispersion only considering the $P_{\rm turb} = \rho v_z^2$ term for each component. Given that the neutral medium (especially, warmer component \UNM+\WNM) dominates the mass fraction (\autoref{tbl:filling}), $\sigma_{\rm eff,\twop}$ agrees well with the effective velocity dispersion of all warm and cold gas at $T<3.5\times10^4\Kel$ presented in \autoref{tbl:satprop}. On the one hand, this makes the observed \ion{H}{1} velocity dispersion a good tracer for the (thermal plus turbulent) velocity dispersion of the mass-dominating component. On the other hand, it shows that \WIM{} tracers will typically overestimate the mass-weighted velocity dispersion.  This could lead to a bias if, for example, H$\alpha$ velocities are used in estimators for the ISM weight (see \autoref{sec:prfm}).  It is also noteworthy that the turbulent velocity dispersion is much lower than the effective velocity dispersion; thermal and magnetic components contribute significantly to the total pressure.

\begin{figure*}
    \centering
    \includegraphics[width=\textwidth]{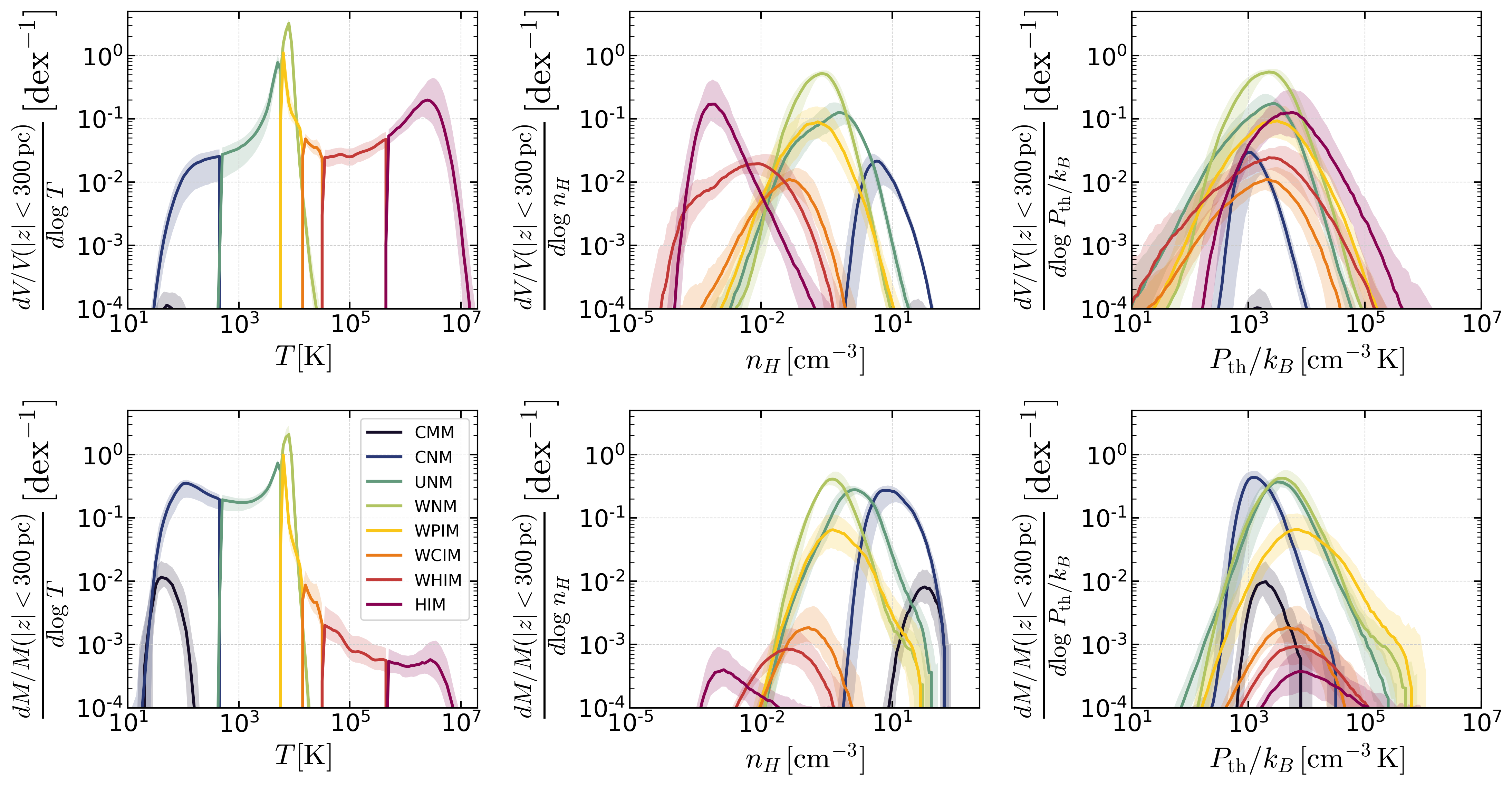}
    \caption{PDFs, separated by phase, of temperature (left), density (middle), and pressure (right) at $|z|<300\pc$ for {\tt R8-4pc}.  Top row shows volume-weighted and bottom shows mass-weighted distributions. The lines show the median over $250\Myr<t<450\Myr$. The shaded regions represent the 16$^{\rm th}$ to 84$^{\rm th}$ percentile range. \label{fig:1D-pdf-R8}}
\end{figure*}

\begin{figure*}
    \centering
    \includegraphics[width=\textwidth]{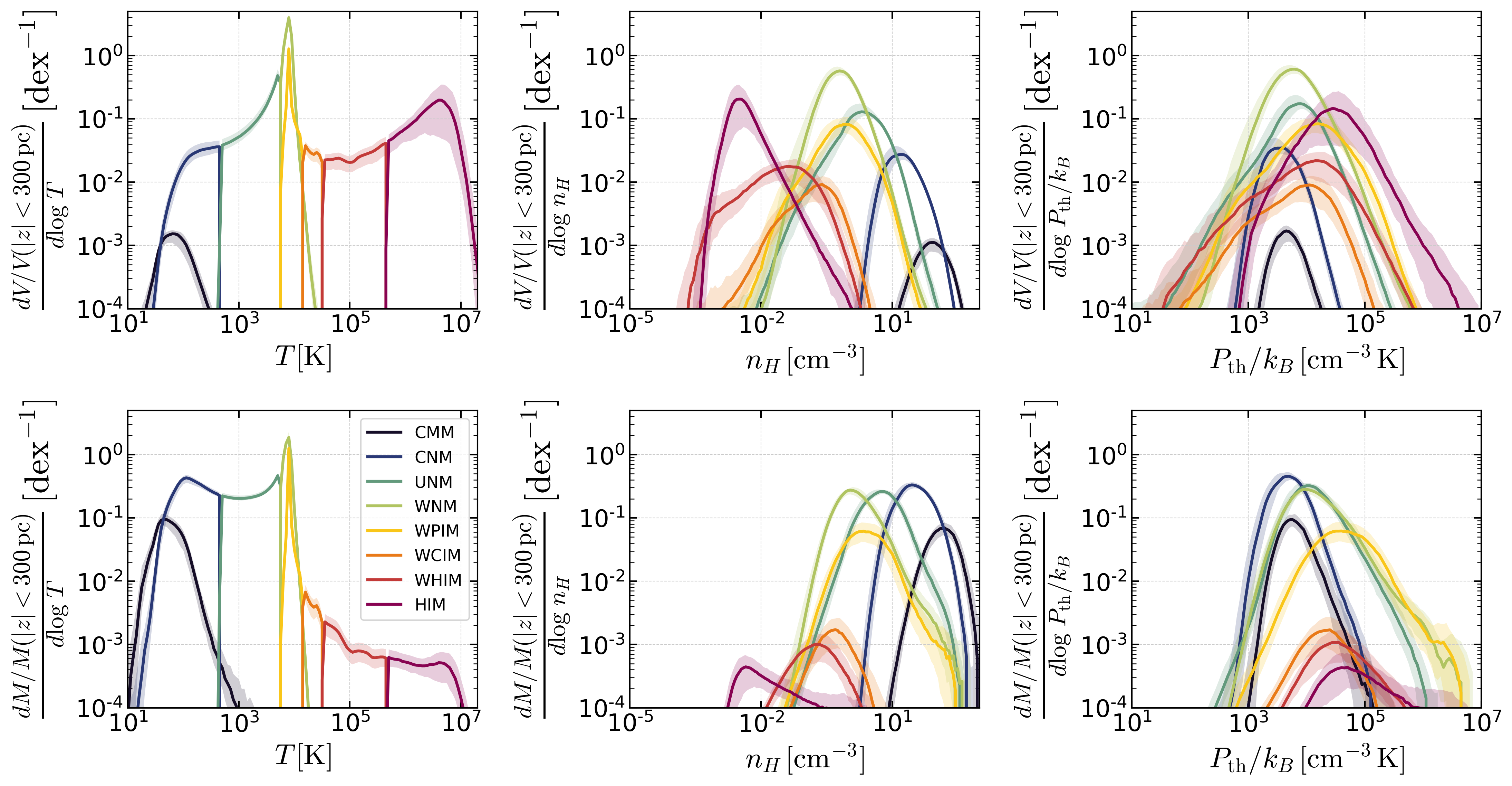}
    \caption{
    Similar to \autoref{fig:1D-pdf-R8}, but for
    \label{fig:1D-pdf-LGR4} {\tt LGR4-2pc} over $250\Myr<t<350\Myr$.}
\end{figure*}

\autoref{fig:1D-pdf-R8} and \autoref{fig:1D-pdf-LGR4} show probability distribution functions (PDFs) of temperature, density, and thermal pressure from {\tt R8-4pc} and {\tt LGR4-2pc}, respectively, based on the region within $|z|<300\pc$. \WNM{} and \WPIM{} have similar characteristic densities and temperatures, but the thermal pressure of \WPIM{} is higher because of the contribution from free electrons. \CMM{} corresponds to the dense part of \CNM{} with similar thermal pressure. All other phase definitions are mostly equivalent to simple temperature cuts. \UNM{} and \WNM{} are in rough thermal pressure equilibrium, but \CNM{} tends to have lower thermal pressure, which is compensated by higher magnetic pressure. \WCIM{} and \WHIM{} are not significant components in terms of mass and volume as they are usually populated only near the interfaces between warm and hot gas (see \autoref{fig:phase}). However, most (net) cooling occurs in these phases (see bottom right panel of \autoref{fig:R8_snapshot}, and \autoref{sec:jointpdf} below). The thermal pressure of \HIM{} is generally larger than that of \WNM{}. Since the thermal pressure of \HIM{} is in balance with the total pressure of \WNM{}, and the turbulent and magnetic contributions in \WNM{} are larger in {\tt LGR4} than in {\tt R8} (see \autoref{sec:prfm}), the difference in thermal pressure between \HIM{} and \WNM{} is larger in {\tt LGR4}.

\subsection{Joint PDF of density and pressure}\label{sec:jointpdf}

\begin{figure*}
    \centering
    \includegraphics[width=\textwidth]{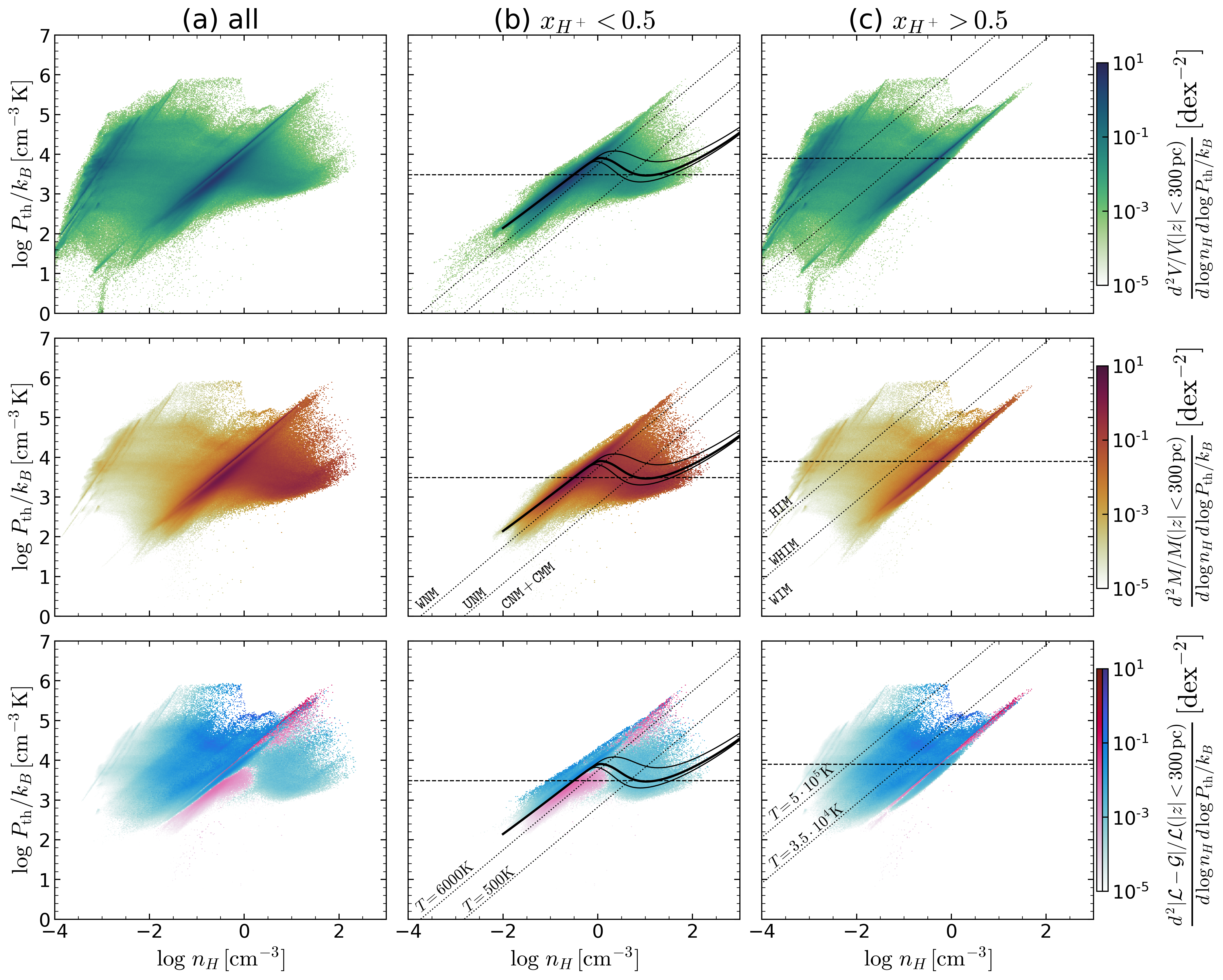}
    \caption{Joint PDFs in $\nH$ and $P_{\rm th}/k_B$ weighted by volume (top), mass (middle), and net cooling rate (bottom) for the gas in {\tt R8-4pc} at $t=280\Myr$ (time corresponds to \autoref{fig:R8_snapshot}(a)). All gas within $|z|<300\pc$ shown in (a) is subdivided into (b) neutral ($\xHII<0.5$) and (c) ionized ($\xHII>0.5$).
    Note that the PDF weighted by net cooling rate is normalized by the total cooling rate within the volume, adopting logarithmic blue-ish and pink-ish color scales for net cooling and heating, respectively. In the middle column, the diagonal dotted lines show $T=500$ and $6000\Kel$ for neutral gas ($P/k_B=1.1 \nH T$).  In the right column, the diagonal dotted lines show $T=3.5\cdot10^4$ and $T=5\cdot10^5\Kel$ for ionized gas ($P/k_B=(1.1+x_e) \nH T$ with $x_e=1$ and 1.2, respectively). The majority of the  \WPIM{} is found near $T\sim7000\Kel$. The neutral medium is distributed somewhat broadly, but with a concentration near $T=7500\Kel$ for \WNM{}, and near the locus corresponding to thermal equilibrium (zero net cooling) for \CNM{}. The \HIM{} region shows tracks roughly following $P\propto \rho^{5/3}$, corresponding to adiabatic expansion of hot bubble interiors. Horizontal dashed lines in (b) and (c) show volume-weighted mean pressure of the neutral gas and ionized gas, respectively.
    In the middle column (b), we show unshielded equilibrium curves for $\xicr=2.9\times10^{-16}{\rm\,s^{-1}}$ and three values of FUV radiation field $\chiFUV=0.51$, 0.87, and 2.6, corresponding to 2$^{\rm nd}$, 50$^{\rm th}$, and 98$^{\rm th}$ percentiles of the volume-weighted $\chiFUV$ distribution within $|z|<300\pc$. The median curve describes the \WNM{} branch well.
    \label{fig:R8_nP}}
\end{figure*}

\autoref{fig:R8_nP} shows, for {\tt R8-4pc} at $t=320\Myr$ within $|z|<300\pc$, the instantaneous distribution of gas in the density-pressure phase plane weighted by volume, mass, and net cooling rate. The total gas is shown in column (a), while column (b) shows the neutral (atomic + molecular) gas and (c) shows the ionized gas, using a cut $\xHII=0.5$. Note that the $x$-axis is the hydrogen number density $\nH$ rather than the total number density $n=(1.1+x_e-0.5\xHH)\nH$. Therefore, at a given temperature the neutral and ionized medium lie on different pressure tracks as a function of $\nH$---a higher (lower) pressure track for the warm ionized (neutral) gas.

In TIGRESS-NCR, the heating and cooling rates are not solely a function of density and temperature and a spatially-uniform FUV field (which was the case in TIGRESS-classic), but also depend on other quantities such as the electron abundance and spatially-nonuniform radiation (see \autoref{eq:Gamma_def} and \autoref{eq:Lambda_def}). Thus, a single thermal equilibrium curve applicable for the whole simulation domain cannot be drawn in \autoref{fig:R8_nP}. Yet, we still see the characteristic locus of thermal equilibrium for neutral gas \citep[see e.g.,][]{1969ApJ...155L.149F,1995ApJ...443..152W,KGKO} in the bottom panel of \autoref{fig:R8_nP}(b) as the boundary between cooling-dominated and heating-dominated regions.
Given $\xicr=2.9\times10^{-16}{\rm\,s^{-1}}$ for the background gas and the median value of $\chiFUV=0.87$, in \autoref{fig:R8_nP}(b), we plot an equilibrium curve as thick solid line (as well as two thin lines for $\chiFUV=0.51$ and 2.6). Since we ignore shielding of FUV in these one zone models, the unshielded equilibrium curves give overall higher pressure at high densities, although the \WNM{} equilibrium branch and its maximum pressure is well described by the median equilibrium curve.

The volume-weighted mean pressure (within $|z|<300\pc$) of the neutral gas, $P/k_B=3.1\times10^3\pcc\Kel$, is shown as a horizontal dashed line. This pressure sits nicely between the maximum thermal equilibrium pressure of \WNM{} (the bulk net heating region shown as pink) at $\nH\sim 0.5\pcc$ and $P/k_B\sim 5\times 10^3\pcc\Kel$ and the minimum thermal equilibrium pressure of the \CNM{} (the bulk net cooling region shown as blue) at $\nH\sim 5\pcc$ and $P/k_B\sim 1\times 10^3\pcc\Kel$.
Although the ionized gas has a very wide range of thermal pressure, the mean is $P/k_B=7.9\times10^3\pcc\Kel$, which is shown as the horizontal dashed line in \autoref{fig:R8_nP}(c). This is higher than that of the neutral gas, in which turbulence and magnetic field significantly contribute to the total pressure (see \autoref{sec:pe_vde}).

As shown in \autoref{fig:1D-pdf-R8}, both mass and volume are dominated by the neutral medium near the disk midplane.
The ionized gas occupies $\sim30-40\%$ by volume (approximately equally in \WIM{} and \HIM) and $\sim10\%$ by mass (mostly in \WIM).
The bottom row of \autoref{fig:R8_nP} shows that both neutral and ionized gas populate a wide parameter space with net cooling or heating (i.e, gas is out of thermal equilibrium). Photoionization can cause net heating in expanding \ion{H}{2} regions as well as the diffuse \WIM{}, which is evident in the narrow dark pink strip at $T\sim7000\Kel$ in the bottom-right panel of \autoref{fig:R8_nP}.
Out-of-equilibrium \CNM{} is mostly in the net (radiative) cooling regime, implying that dissipation of turbulence may contribute to the thermal balance in \CNM{} to allow an overall excess of radiative cooling over radiative heating. Locally, \WNM{} is perturbed into both net cooling and heating regimes, while \WNM{} as a whole is experiencing net cooling.

\begin{figure*}
    \centering
    \includegraphics[width=\linewidth]{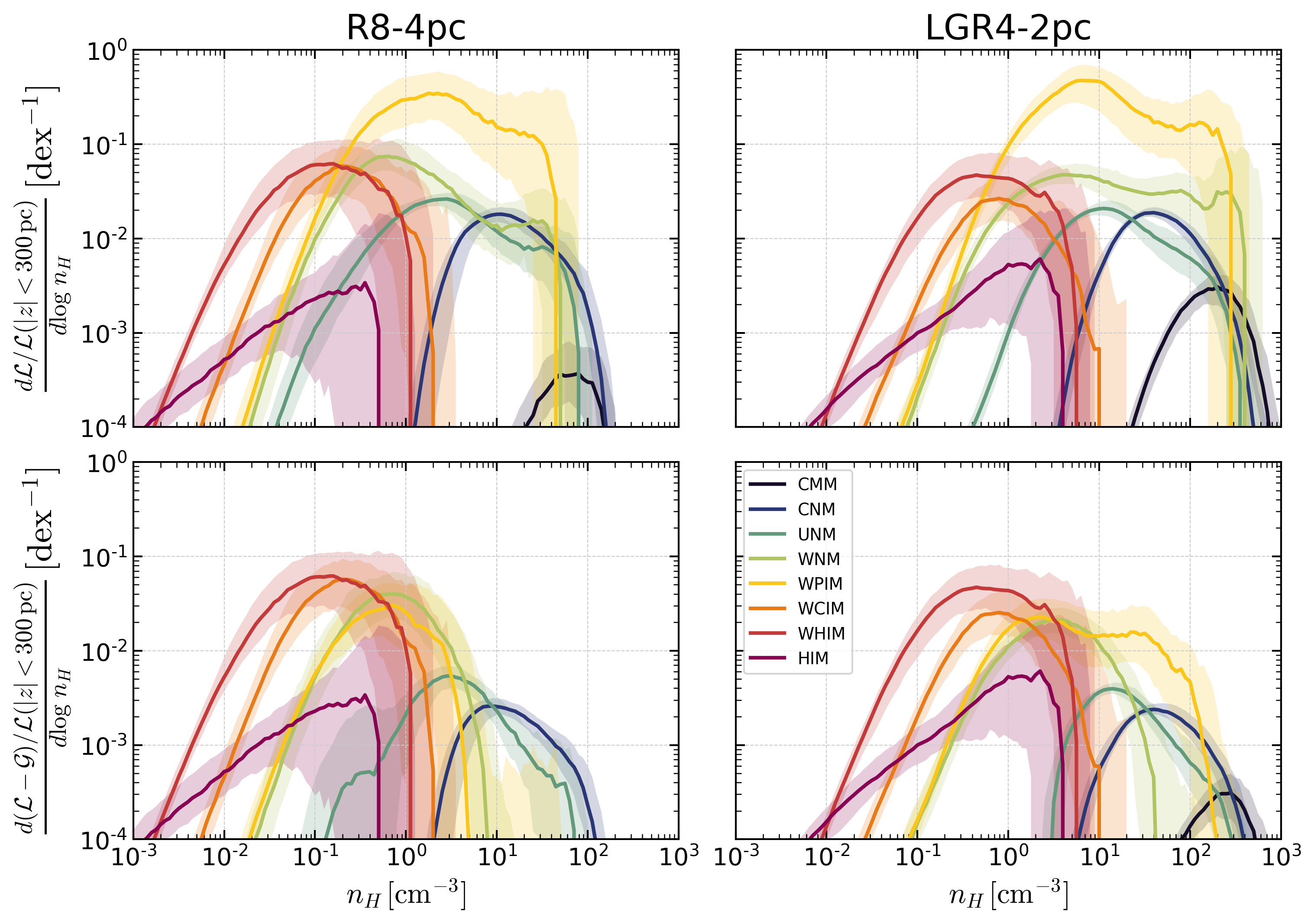}
    \caption{Distribution in $\nH$ of gas within $|z|<300\pc$, separated by phase and weighted by the cooling rate (top) and net cooling rate (bottom) for {\tt R8-4pc} (left) and {\tt LGR4-2pc} (right).  Overall normalization is by the total cooling rate within the volume. The lines show the median over $250\Myr<t<450\Myr$ for {\tt R8-4pc} and  $250\Myr<t<350\Myr$ for {\tt LGR4-2pc}. The shaded regions represent the 16$^{\rm th}$ to 84$^{\rm th}$ percentile range. Total cooling is by far dominated by \WPIM{} ($\sim 70\%$), while net cooling is greatest in \WHIM{}.}
    \label{fig:pdf-cooling}
\end{figure*}

Due to its low density, cooling in the hot gas located inside SN(e)-driven bubbles is negligible. The corresponding high-temperature regions in \autoref{fig:R8_nP} show adiabatic expansion tracks, $P\propto \rho^{5/3}$. The thermalized SN energy is mostly cooled away at lower temperature phases, e.g., \WHIM{} and \WNM/\WIM. To clarify the contribution of each gas phase in cooling,
\autoref{fig:pdf-cooling} plots the cooling and net cooling contribution within $|z|<300\pc$ from each phase as a function of density, for {\tt R8-4pc} (left) and {\tt LGR4-2pc} (right).
The total radiative cooling rate per volume, shown in the top panel, is dominated by \WPIM{} ($\sim70\%$). \WNM{} and \WHIM{} contribute about 10\% each. But, \WPIM{} and \WNM{} are also the gas phases within which most radiative (photoionization and photoelectric) heating occurs. The bottom panels show the net cooling rate per volume, with heating subtracted from cooling. The net cooling, which when integrated over volume produces the history shown in \autoref{fig:tevol_ss}(c), is now dominated by \WHIM{} ($\sim 40\%$). \WNM, \WPIM, and \WCIM{} contribute about 10-20\% each.

\section{Pressure-Regulated, Feedback-Modulated Theory of the Equilibrium Star-Forming ISM}\label{sec:prfm}

Having presented the overall characteristics of the simulated ISM, in this section we focus on the midplane pressure and stresses, gas weight, and SFR surface density, and their mutual correlations. These analyses aim to confirm the validity and prediction of the 
PRFM 
star formation theory, first formulated in \citet{2010ApJ...721..975O} and \citet{2011ApJ...731...41O} and tested in subsequent work \citep{2011ApJ...743...25K,2013ApJ...776....1K,2012ApJ...754....2S,2015ApJ...815...67K}.
\REV{}{For a comprenhensive summary and detailed derivation of the theory, the reader is referred to \citet{2022ApJ...936..137O}.} We closely follow the analysis of \citet{2022ApJ...936..137O}, which \REV{summarizes the theory and}{} analyzes the TIGRESS-classic simulation suite.

\REV{Here, we simplify our full phase definition into three phases: \twop{} for the neutral medium at cold to warm temperatures (i.e., \twop=\CMM+\CNM+\UNM+\WNM), \WIM{} for the warm ionized medium (i.e., \WIM=\WPIM+\WCIM), and \hot{} for the hot medium (i.e., \hot=\WHIM+\HIM).
Note that in Ostriker \& Kim (2022), we only had \twop{} and \hot{} as the TIGRESS-classic framework does not follow the ionization state explicitly.
}{}

The PRFM theory views the ISM in galactic disks as a long-lived thermal-dynamical system with stellar feedback as the main energy source. Despite the vigorous dynamical evolution seen in our simulations (and in the real ISM), the system is in a quasi-steady state on average (in terms either of long-term temporal averages or large-scale ensemble averages). Under this assumption, the governing gas dynamics equations dictate vertical dynamical equilibrium, a balance between total pressure and gas weight. The ISM energy would drop quickly through cooling and dissipation in the absence of inputs, but stellar feedback can offset losses to maintain the required pressure/stress. As a consequence, the PRFM theory posits that galactic SFRs are naturally linked to the dynamical equilibrium pressure, which in turn predicts galactic SFRs from large-scale mean galactic parameters.

\REV{}{Numerical simulations that directly capture self-consistent energy injection (by feedback) and 
loss processes are critical in determining the feedback yield. In the TIGRESS-NCR framework, we do not impose radiation fields and the resulting photoheating based on
observational estimates, but compute $J_{\rm FUV}$ via ray-tracing from young star cluster particles formed in our simulations, where the number and location of these star particles is self-consistently set by the rate of gravitational collapse. Similarly, our simulations have sufficient resolution to follow the transition from adiabatic to cooling stages of SN remnant expansion. Thus, unlike in lower resolution simulations, we resolve the simultaneous heating and acceleration of ambient gas by SN shocks as well as subsequent cooling.
The present simulations solving direct UV radiation transfer not only for FUV but also for EUV are critical for validation of the simpler treatment for FUV heating used in TIGRESS-classic, and for accurately calibrating the feedback yields.
}

Our analysis steps in this section are as follows. We first check pressure equilibrium among the three phases and the vertical dynamical equilibrium between total pressure support and gas weight (\autoref{sec:pe_vde}). Then, we measure each pressure component and examine the pressure-$\Ssfr$ relation (\autoref{sec:yields}). This gives a numerical calibration of the key parameter in the theory, the ratio of pressure and $\Ssfr$, which we call the \emph{feedback yield}. We compare our new results for the feedback yield with the theoretical and numerical results in \citet{2022ApJ...936..137O}. Since we only consider two galactic conditions in this paper, we refrain from deriving new fitting results. In \autoref{sec:P_SFR}, we present the correlations between SFR surface density, total pressure, and weight (or its simplified estimator, dynamical equilibrium pressure).
\REV{Crucially, in TIGRESS-NCR, we do not impose radiation fields based on (scaled) observational estimates, but compute $J_{\rm FUV}$ using radiative transfer from young star cluster particles formed in our simulations, where the number and location of these star particles is self-consistently set by the rate of gravitational collapse. Similarly, the present simulations have sufficient resolution to follow the transition from adiabatic to cooling stages of SN remnant expansion.  Thus, unlike in lower resolution simulations, we resolve the simultaneous heating and acceleration of ambient gas by SN shocks as well as subsequent cooling.  Our simulations therefore directly capture both the energy injection and energy dissipation processes that enter in determining the feedback yield.}{}

\subsection{Pressure Equilibrium and Vertical Dynamical Equilibrium}\label{sec:pe_vde}

\begin{figure*}
    \centering
    \includegraphics[width=\textwidth]{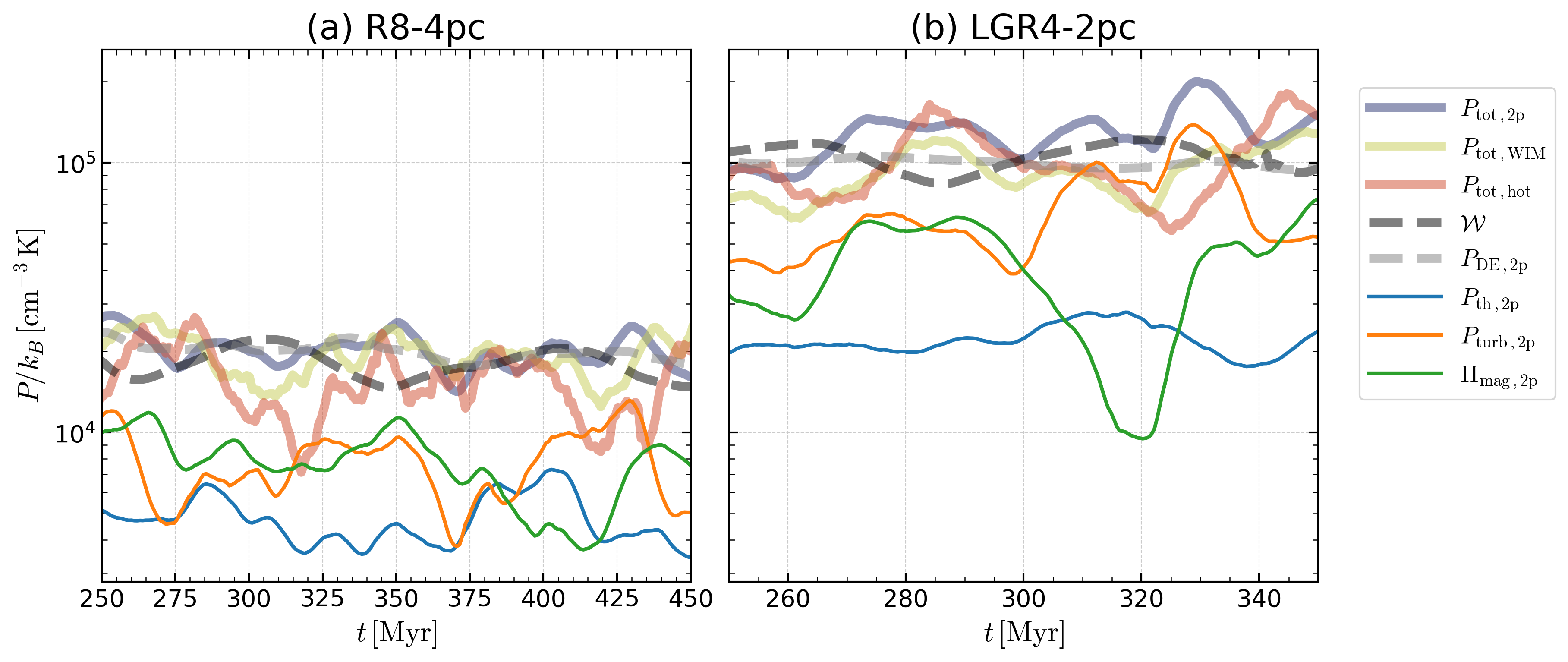}
    \caption{Time evolution of midplane pressures and weight. Total midplane pressure of \twop{} (black), \WIM{} (yellow), and \hot{} (red) phases are comparable with each other. This total vertical support matches  the ISM weight (black dashed, dominated by \twop). The simple weight estimator $\PDE$ (grey dashed) provides reasonable agreement with the weight and total pressure. We show each pressure component of \twop{} as blue ($\Pthtwo$), orange ($\Pturbtwo$), and green ($\Pimagtwo$) lines.
    \label{fig:Pmid_tevol}}
\end{figure*}

\REV{The PRFM formulation assumes that temporally and horizontally averaged vertical momentum and energy equations
(Equation 2 and Equation 3)
satisfy a steady state.}{}
By integrating the vertical
\REV{}{component of the} momentum equation
\REV{}{(\autoref{eq:mom_con})}
from the midplane to the top/bottom of the gas disk, the vertical dynamical equilibrium condition \REV{}{assuming a steady state} is then given by the balance between the midplane total pressure \REV{}{($\Ptot$)} and the ISM weight \REV{}{($\W$)}:
\begin{equation}\label{eq:zmom_avg}
    \Delta \abrackets{\Ptot} \equiv
    \Delta\abrackets{\Pth + \Pturb + \Pimag} + \Delta P_{\rm rad} = \W
\end{equation}
where $\Delta$ denotes the difference between the midplane ($z=0$) and the top/bottom of the gaseous disk ($z=\pm L_z/2$), and the angle brackets denote a horizontal average.
In \REV{the above}{\autoref{eq:zmom_avg}},
we adopt the following nomenclature of pressure components: thermal pressure $\Pth=P$, turbulent pressure (Reynolds stress) $\Pturb\equiv\rho v_z^2$, and magnetic stress (magnetic pressure + tension) $\Pimag \equiv (B_x^2+B_y^2-B_z^2)/(8\pi)$. Note that the mean or turbulent magnetic stress ($\oPimag$ or $\dPimag$) is respectively defined by substituting for $\Bvec$ with the mean $\overline{\Bvec} \equiv \abrackets{\Bvec}$ or turbulent $\delta{\Bvec}\equiv \Bvec-\overline{\Bvec}$ component.
The radiation pressure and weight terms can be defined toward either upper or lower disk, $\Delta P_{\rm rad} = \int_0^{\pm L_z/2}\abrackets{f_{\rm rad,z}}dz$ and $\W = \int_0^{\pm L_z/2}\abrackets{\rho \partial{\Phi}/\partial{z}}dz$. We take an average of two values (integrated from top or bottom) in the following analysis.
The vertical gravity $-\partial \Phi/\partial z$ is a sum of terms from stars, dark matter, and gas, so the total weight can be decomposed into two terms: gas weight from  external gravity  $\W_{\rm ext}$ (due to stellar disk and dark matter halo), and gas weight from self-gravity $\W_{\rm sg}$ (due to gas).
Generally, the pressure components at the midplane $z=0$ are much larger than those at the top of the gaseous disk, leading to $\Delta P\rightarrow P(z=0)$.

To separate the contribution from each phase, we define the horizontal average of a quantity $q$ for a selected phase by
\begin{equation}\label{eq:midavg}
    \abrackets{q}_{\rm ph}\equiv \frac{\int\int q\Theta({\rm ph}) dxdy}{L_x L_y}
\end{equation}
where $\Theta({\rm ph}) = 1$ if temperature and abundance condition in  \autoref{tbl:phase} is satisfied and 0 otherwise.
\REV{for ph=\twop, \WIM, and \hot.}{Here, we simplify our full phase definition into three phases: \twop{} for the neutral medium at cold to warm temperatures (i.e., \twop=\CMM+\CNM+\UNM+\WNM), \WIM{} for the warm ionized medium (i.e., \WIM=\WPIM+\WCIM), and \hot{} for the hot ionized medium (i.e., \hot=\WHIM+\HIM).
}
We can also separately define the area filling factor $f_{\rm A,ph} \equiv \int\int\Theta({\rm ph})dxdy/L_xL_y$.

Each phase's contribution adds up such that $\abrackets{\Ptot} = \sum_{\rm ph} \abrackets{\Ptot}_{\rm ph}$.  Individual pressure components ($\abrackets{P_{\rm th}}$, etc.) can be written as a sum over contributions from each phase in the same way.
The typical midplane value of the total pressure in each phase is defined by $\tilde{P}_{\rm tot,ph}\equiv\abrackets{\Ptot}_{\rm ph}/f_{\rm A, ph}$ (and similarly for $\tilde{P}_{\rm th, ph}$ etc.). We can then write $\abrackets{\Ptot} =\sum_{\rm ph} f_{\rm A, ph} \tilde{P}_{\rm tot,ph}$.
If the typical values of the total pressure for \twop, \WIM, and \hot{} are comparable with each other,
we have $\abrackets{\Ptot} \approx \tilde{P}_{\rm tot, X}\sum_{\rm ph} f_{\rm A,ph} = \tilde{P}_{\rm tot, X}$, where X denotes any given phase.
If we neglect the direct UV radiation pressure $\Delta P_{\rm rad}$ for succinctness as it contributes less than 1\% to the total pressure in both models,
\autoref{eq:zmom_avg} simply becomes
\begin{equation}\label{eq:zmom_avg_2p}
\Delta\abrackets{\Ptot} \approx \tilde{P}_{\rm tot,2p} = \tilde{P}_{\rm th,2p} + \tilde{P}_{\rm turb,2p} + \tilde{\Pi}_{\rm mag,2p} = \W.
\end{equation}
We note that weight (and radiation pressure) is vertically integrated over all phases, and that the (time-averaged) pressure of gas in any given phase at the midplane must support the weight of gas in all phases above it (rather than selectively supporting its own phase).
In our simulations, the gas weight is dominated by \twop{} with 14\% (4\%) from \WIM{} for {\tt R8} ({\tt LGR4}) and less than 1\% from \hot. The contribution from the external gravity is 75\% for {\tt R8} and 30\% for {\tt LGR4}.

\autoref{fig:Pmid_tevol} shows time evolution of all pressure terms in \autoref{eq:zmom_avg_2p} along with the total midplane pressures of \WIM{} and \hot{}. In this and other figures and tables, the values of pressures shown represent midplane averages either within a given phase or over all phases, dropping the tilde for cleaner notation. Comparing the total pressures of each phase (dark blue for \twop, yellow for \WIM, and red for \hot), we confirm that they are roughly in pressure equilibrium.
Also shown are the direct measure of the ISM weight ($\mathcal{W}$) and the commonly used weight estimator \REV{}{(which assumes that the stellar disk is thicker than the gaseous disk; \citealt{2022ApJ...936..137O})}
\begin{equation}\label{eq:PDE}
\PDEtwo \equiv \frac{\pi G \Sgas^2}{2} + \Sgas (2G\rho_{\rm sd})^{1/2} \sigma_{\rm eff, 2p},
\end{equation}
constructed from observables \citep[e.g.,][]{2020ApJ...892..148S}.
We note that $\sigma_{\rm eff,\twop}$ presented in \autoref{tbl:veld} is the mass-weighted mean for the \twop{} phase over the entire domain (not the midplane measure).
This kinetic (thermal+turbulent) velocity dispersion is a direct observable given line emission from the neutral (atomic and molecular) gas, and then $\sigma_{\rm eff,\twop}$ can be obtained by correcting for the magnetic terms.
On average, the total pressure and weight are in good agreement with each other (they are usually off-phased).

\begin{deluxetable*}{lCCCCCCCCC}
\tablecaption{Midplane Pressure and Weight \label{tbl:pressure}}
\tablehead{
\colhead{Model} &
\dcolhead{\Pthtwo} &
\dcolhead{\Pturbtwo} &
\dcolhead{\Pimagtwo} &
\dcolhead{\Ptottwo} &
\dcolhead{\Ptotwim} &
\dcolhead{\Ptothot} &
\dcolhead{\Ptot} &
\dcolhead{\W} &
\dcolhead{\PDEtwo}
}
\startdata
{\tt R8-4pc    } & 4.6^{+1.5}_{-0.8} & 7.4^{+2.5}_{-1.9} & 7.7^{+2.4}_{-2.7} & 20.2^{+2.9}_{-2.5} & 19.4^{+3.7}_{-3.6} & 15.1^{+5.2}_{-3.8} & 18.5^{+3.1}_{-3.4} & 17.6^{+2.8}_{-1.9} & 20.1^{+1.2}_{-1.5} \\
{\tt R8-8pc    } & 4.3^{+3.5}_{-1.2} & 7.0^{+5.1}_{-2.0} & 8.5^{+3.0}_{-2.5} & 20.6^{+6.7}_{-4.8} & 20.5^{+6.1}_{-5.4} & 17.8^{+6.2}_{-6.3} & 20.5^{+5.7}_{-5.7} & 19.1^{+5.6}_{-2.0} & 21.5^{+2.7}_{-2.4} \\
{\tt LGR4-2pc  } & 2.1^{+0.4}_{-0.1} & 5.6^{+3.6}_{-1.3} & 4.4^{+1.4}_{-2.2} & 13.0^{+1.5}_{-2.6} & 9.1^{+2.5}_{-1.9} & 9.5^{+4.5}_{-2.1} & 10.5^{+2.8}_{-2.2} & 10.7^{+1.1}_{-1.4} & 10.0^{+0.3}_{-0.4} \\
{\tt LGR4-4pc  } & 2.1^{+0.9}_{-0.7} & 6.2^{+1.5}_{-2.9} & 5.2^{+2.9}_{-3.1} & 12.4^{+5.6}_{-2.4} & 9.0^{+2.7}_{-1.2} & 8.5^{+2.9}_{-1.5} & 10.4^{+3.8}_{-2.5} & 10.9^{+1.5}_{-2.0} & 9.9^{+0.5}_{-0.7} \\
\enddata
\tablecomments{
Each column shows the median and 16$^{\rm th}$ and 84$^{\rm th}$ percentile range over $t=250-450\Myr$ and $t=250-350\Myr$ for {\tt R8} and {\tt LGR4}, respectively. Pressure/weight is in units of $k_B\Kel\pcc$, with a multiplication factor of $10^3$ and $10^4$ for {\tt R8} and {\tt LGR4}, respectively.
}
\end{deluxetable*}

\begin{figure*}
    \centering
    \includegraphics[width=\textwidth]{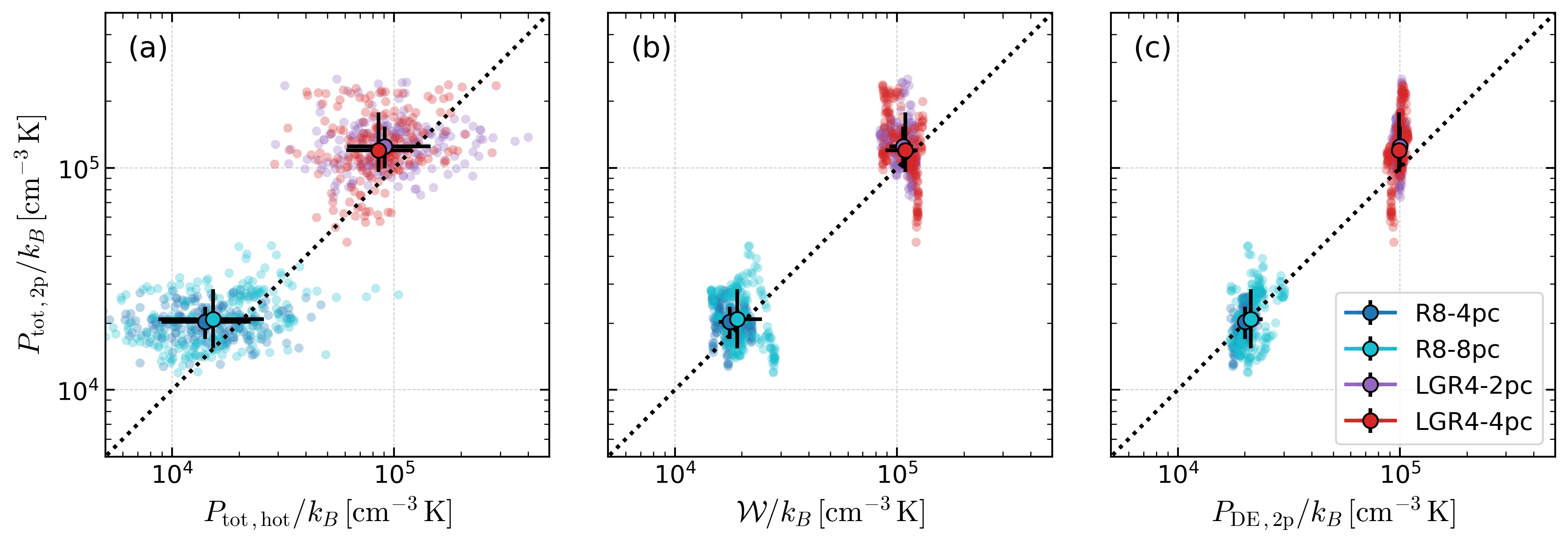}
    \caption{Measured midplane total pressure of warm-cold (\twop) gas $\Ptottwo$ as a function of (a) measured midplane total pressure of the \hot{} phase $\Ptothot$, (b) measured weight $\W$, and
    (c) dynamical equilibrium pressure $\PDEtwo$ (simple weight estimator).
    Individual points at intervals 1 (0.5) Myr are plotted for {\tt R8} ({\tt LGR4}) over 200 (100) Myr interval. Medians with 16$^{th}$ and 84$^{th}$ percentiles are shown as a larger point with errorbars. For reference, the dotted line shows the identity relation.
    \label{fig:vequil}}
\end{figure*}

\autoref{fig:vequil} plots $\Ptottwo$ as a function of (a) $\Ptothot$, (b) $\W$, and (c) $\PDEtwo$, while
\autoref{tbl:pressure} summarizes the midplane pressure components in \twop{} as well as total pressure in each phase, weight, and weight estimator.
Again, approximate pressure equilibrium among the different phases holds, but \hot{} gives slightly lower total pressure. Thermal pressure dominates in \WIM{} and \hot, while thermal pressure is the smallest component in \twop{}, with magnetic and turbulent components comparable.
$\Ptottwo\approx \W$ directly demonstrates that the ISM pressure is \textit{regulated} in disk systems as it obeys the conservation ``law'' of momentum (on average). \autoref{fig:vequil}(c) demonstrates the validity of $\PDEtwo$ as a reasonable estimator of the true weight (see \autoref{tbl:pressure}) and hence total midplane pressure.

\subsection{Feedback Modulation and Yields}\label{sec:yields}

The PRFM theory postulates that thermal and turbulent pressure ($\propto$ energy density) components are sourced by feedback from massive young stars through heating by UV radiation and turbulence driven by SNe. The balance between radiative cooling and heating sets the thermal pressure, while the balance between turbulence driving and dissipation sets the turbulent pressure. Magnetic fields are set by the saturation of galactic dynamo, providing the contribution from the magnetic term at a level similar to (or slightly below) the turbulent term \citep{2015ApJ...815...67K}. The pressure components are thus expected to scale with the rate of feedback energy injection, and therefore with $\Ssfr$. We define the feedback yields $\Upsilon_c\equiv P_c/\Ssfr$ as the ratios of a pressure component $c$ to $\Ssfr$, to quantify the feedback modulation of individual pressure components. Note that the natural unit for the feedback yield is velocity.

\subsubsection{Thermal Pressure}\label{sec:th_yield}
\begin{figure*}
    \centering
    \includegraphics[width=\linewidth]{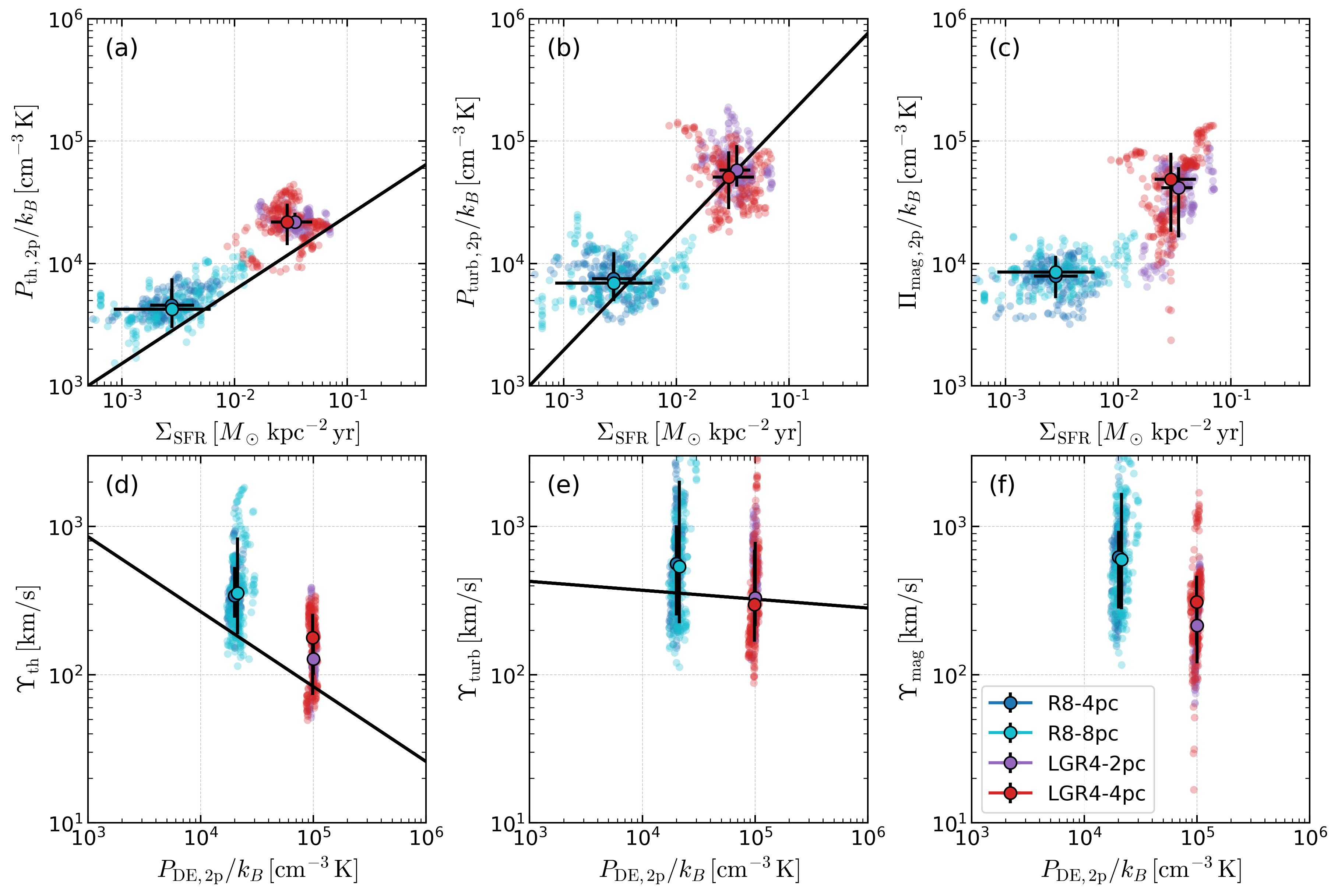}
    \caption{{\bf Top:} Midplane (a) thermal pressure $\Pthtwo$, (b) turbulent $\Pturbtwo$, and (c) magnetic stress $\Pimagtwo$ of the \twop{} medium as a function of SFR surface density $\Ssfr$.
    {\bf Bottom:} Feedback yield for (d) thermal $\Upsilon_{\rm th}\equiv \Pthtwo/\Ssfr$ (e) turbulent $\Upsilon_{\rm turb}\equiv \Pturbtwo/\Ssfr$, and (f) magnetic $\Upsilon_{\rm mag}\equiv \Pimagtwo/\Ssfr$ component as a function of $\PDEtwo$. Individual points at intervals 1 (0.5) Myr are plotted for {\tt R8} ({\tt LGR4}) over 200 (100) Myr interval. Medians with 16$^{th}$ and 84$^{th}$ percentiles are shown as a larger point with errorbars. For reference, the solid lines show the best fit results from \citet{2022ApJ...936..137O}: (a) \autoref{eq:Pth_tigress}, (b) \autoref{eq:Ptrb_tigress}, (d) \autoref{eq:Yth_tigress}, and (e) \autoref{eq:Ytrb_tigress}.}
    \label{fig:P_Y}
\end{figure*}

Because of the short cooling time in the cold and warm ISM (\twop{} and \WIM), the energy gains from radiative heating are quickly lost through optically thin radiative cooling mostly within the same phase.
Thermal pressure in both \twop{} and \WIM{} is then expected to scale with the radiative heating rate, which is proportional to the mean UV intensity and hence SFR.
In the \twop{} medium, the photoelectric effect by FUV is the main heating source.
Therefore, $\Pthtwo\propto \Gamma_{\rm PE}\propto \epsilon_{\rm PE} J_{\rm FUV}$, where $\epsilon_{\rm PE}$ is the photoelectric heating efficiency, which depends sensitively on the grain size distribution and the grain charging parameter $\psi \equiv G_0\sqrt{T}/n_e$ \citep[e.g.,][]{1994ApJ...427..822B,2001ApJS..134..263W}. The source of FUV radiation is massive young stars (the luminosity-weighted mean age of FUV emitters is $\sim10\Myr$) so that $J_{\rm FUV}= \ftau\dot{S}_{\rm FUV}/(4\pi)$ for $\ftau$ a factor accounting for UV radiative transfer in the dusty ISM, where $\dot{S}_{\rm FUV}=L_{\rm FUV}/L_xL_y \propto \Ssfr$.\footnote{We note that we have changed notation for the FUV luminosity per area from $\Sigma_{\rm FUV}$ (e.g., \citealt{2010ApJ...721..975O,2022ApJ...936..137O}) to $\dot{S}_{\rm FUV}$ to consistently refer to areal energy gain and loss rates using $\dot{S}$ (see \autoref{sec:energetics})}.

A simple radiation transfer solution for uniformly-distributed sources in a uniform slab gives
\begin{equation}\label{eq:ftau}
    \ftau = \frac{1-E_2(\tau_\perp/2)}{\tau_\perp}
\end{equation}
\citep{2010ApJ...721..975O}.
Here, $E_2$ is the second exponential integral and $\tau_\perp=\kappa_{\rm FUV}\Sigma_{\rm gas}$ is the mean optical depth to FUV. For $\kappa_{\rm FUV}=10^3\cm^2\gram^{-1}$, $\ftau\approx 1/\tau_\perp$ at $\Sgas>20\Surf$.
In the TIGRESS-classic suite, we adopted the approximate form of $\ftau$ as presented in \autoref{eq:ftau} to convert $\dot{S}_{\rm FUV}$ to $J_{\rm FUV}$, and we also adopted a single value for $\epsilon_{\rm PE}$.

The attenuation of FUV increases at higher surface densities (which generally corresponds to higher pressures). The relationship between the thermal pressure and $\Ssfr$ is thus sublinear, resulting in a decrease of thermal pressure yield at higher $\PDE$.
The fit to the TIGRESS-classic suite gives \citep{2022ApJ...936..137O}
\begin{subequations}
\begin{eqnarray}
    \label{eq:Pth_tigress}
    \log(\Pthtwo/k_B) = 0.603\,\log(\Ssfr) &+& 4.99\\
    \label{eq:Yth_tigress}
    \log \Upsilon_{\rm th} =-0.506\,\log\left({\PDE}/{k_B}\right) &+& 4.45.
\end{eqnarray}
\end{subequations}

In the current simulations, the connection from $\Sigma_{\rm SFR}$ to $J_{\rm FUV}$ to $\Gamma_{\rm PE}$ is self-consistently determined by explicit UV radiation transfer and an adopted theoretical dust model for the heating efficiency \REV{ (our standard choice is a model with grain size distribution A, $R_V=3.1$, and $b_C=4.0\times10^{-5}$ in (Weingartner \& Draine 2001b).}{and cross sections.}
\autoref{fig:P_Y}(a) plots $\Pthtwo$ vs. $\Ssfr$, showing the similar sublinear relationship calibrated to TIGRESS-classic (\autoref{eq:Pth_tigress}; solid black line).  \autoref{fig:P_Y}(d) shows the thermal feedback yield that
decreases as $\PDE$ increases (the TIGRESS-classic fit \autoref{eq:Yth_tigress} is also shown). We find that the scaling is quite similar to the fit from the TIGRESS-classic suite, but the normalization in $\Upsilon_{\rm th}$ is higher here -- that is, the TIGRESS-NCR simulations give rise to the higher thermal pressure at a given $\Ssfr$ than the TIGRESS-classic simulations. The offset is because the explicit treatment of the heating efficiency here yields on average a factor of 2--3 higher heating rate for a given $J_{\rm FUV}$, compared to the heating rate coefficient adopted in TIGRESS-classic (which is from \citealt{2002ApJ...564L..97K}). The consistent scaling \REV{stems from the fact}{suggests} that \autoref{eq:ftau} is indeed a good approximation for the mean attenuation factor in comparison to the actual radiation transfer solution obtained here by adaptive ray tracing (N. Linzer et al. in prep.).

\subsubsection{Turbulent Pressure}\label{sec:turb_yield}

The turbulent pressure in the warm and cold components of the ISM arises from large-scale forcing, with expanding hot bubbles produced by SN feedback as the most important source.  Because the energy injection from SNe is highly localized in space and time, it creates a shock when it is transferred to the warm and cold ISM gas.  This accelerates the surrounding ISM, increasing the total momentum until the shock becomes radiative when the post-shock temperature $T\lesssim 10^6\Kel$ (or $v_{\rm SNR}\sim 200\kms$).
The radial momentum injected per SN ($p_*$) is much larger ($\sim 10^5\Msun\kms$) than the momentum of the initial SN ejecta ($\sim 10^4\Msun\kms$) because the shock accelerates two orders of magnitude more mass than the initial ejecta before becoming radiative. The SN momentum injection is also much greater than other sources, such as expanding \ion{H}{2} regions \citep{2018ApJ...859...68K} and stellar wind driven bubbles \citep{2021ApJ...914...90L,2021ApJ...914...89L}.

For the \citet{2001MNRAS.322..231K} IMF, the total stellar mass formed for every SN progenitor star is $m_*\sim100\Msun$, and the areal rate of SN explosions in quasi-steady state is $\Ssfr/m_*$.
For spherical momentum injection per SN of $p_*$ centered on the disk midplane,
\citet{2011ApJ...731...41O} argued that
the turbulent pressure $\rho v_z^2$ is expected to be comparable to the rate of vertical momentum flux injected on either side of the disk, $P_{\rm turb} = (p_*/4)(\Ssfr/m_*)$.
Since $p_*$ is insensitive to the environment (both density and metallicity; e.g., \citealt{2015ApJ...802...99K,2017ApJ...834...25K,KGKO}), the turbulent feedback yield is expected to be nearly constant (this is in stark contrast to the thermal feedback yield).
The fit to the TIGRESS-classic suite gives \citep{2022ApJ...936..137O}
\begin{subequations}
\begin{eqnarray}
    \label{eq:Ptrb_tigress}
    \log(\Pturbtwo/k_B) = 0.96\,\log(\Ssfr) &+& 6.17\\
    \label{eq:Ytrb_tigress}
    \log \Upsilon_{\rm turb} =-0.06\,\log({\PDE}/{k_B})&+& 2.81.
\end{eqnarray}
\end{subequations}

\autoref{fig:P_Y}(b) plots  $\Pturb$ vs. $\Ssfr$, showing the expected near linear relationship (\autoref{eq:Ptrb_tigress}). The turbulent feedback yield shown in \autoref{fig:P_Y}(e) is consistent with the shallow dependence on $\PDE$ seen in \autoref{eq:Ytrb_tigress}. We note that the current simulations have additional momentum injection by expanding \ion{H}{2} regions as well as direct UV radiation pressure. Apparently, the contribution of UV in modulating global turbulent pressure is not significant. This strongly contrasts with the dominant role of UV in the destruction of molecular clouds \citep[e.g.,][]{2018ApJ...859...68K,2021ApJ...911..128K}.

\subsubsection{Magnetic Stress}\label{sec:mag_yield}
We find that the midplane magnetic stress and hence magnetic feedback yield  (\autoref{fig:P_Y}(c) and (f)) is comparable to the turbulent kinetic component, for both models.
Magnetic terms are determined by galactic dynamo action as a result of the interaction between turbulence (driven by feedback), galactic differential rotation, and buoyancy. The turbulent component of magnetic fields is directly related to the kinetic energy in  turbulence and turbulent magnetic energy density quickly saturates at a level similar to kinetic energy density as long as the initial field is strong enough \citep{2015ApJ...815...67K}.
Our initial field is purely azimuthal (along the $y$ direction) and comparable to the final saturation level.
Overall, the current simulations cover long-term evolution and result in a saturated state without a sign of further secular evolution in magnetic field strengths.\footnote{
The regular (mean) component of magnetic fields is maintained in our simulations as we include galactic differential rotation using the shearing box. In separate experiments without rotation or weak shear, we find much lower saturation level of magnetic fields and hence magnetic stress. We defer the detailed exploration and discussion of the magnetic field evolution to a separate work.}

\subsection{Total pressure and SFR prediction}\label{sec:P_SFR}

\begin{figure*}
    \centering
    \includegraphics[width=\textwidth]{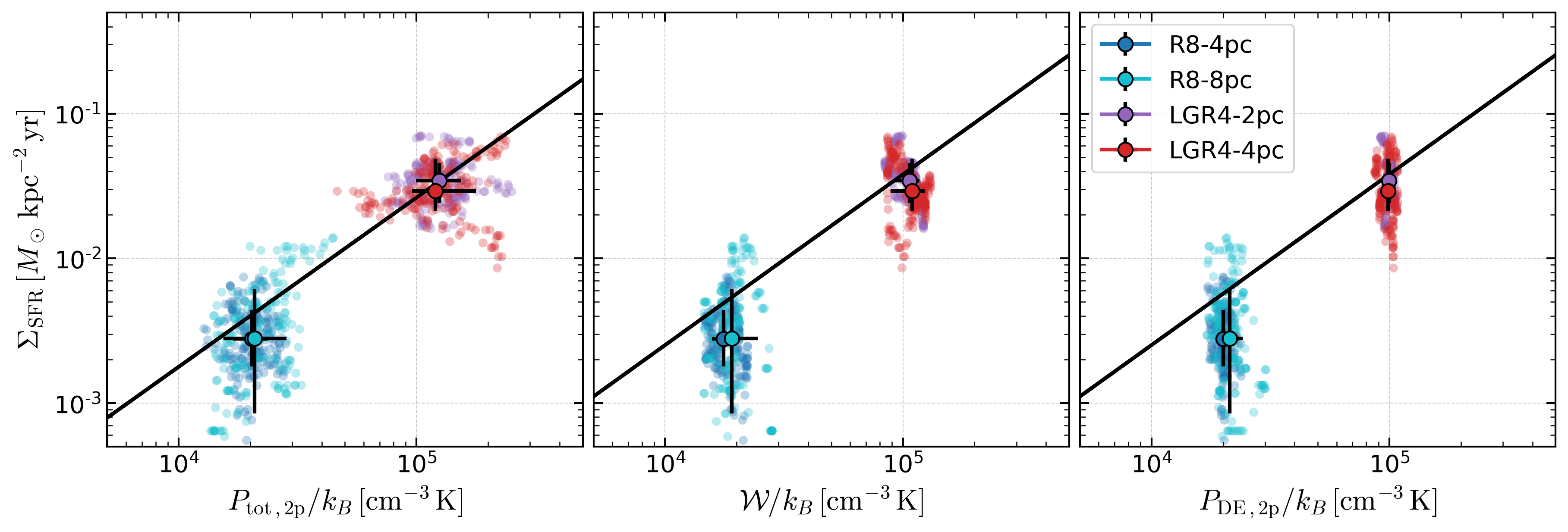}
    \caption{$\Ssfr$ as a function of measured total midplane pressure $\Ptottwo$,
    measured ISM weight  $\W$, and estimated weight $\PDEtwo$. Individual points at intervals 1 (0.5) Myr are plotted for {\tt R8} ({\tt LGR4}) over 200 (100)Myr interval. Medians with 16$^{\rm th}$ and 84$^{\rm th}$ percentiles are shown as a larger point with errorbars. For reference, the solid lines show the best fit results from \citet{2022ApJ...936..137O} (\autoref{eq:SFR_P} to \ref{eq:SFR_PDE}).}
    \label{fig:P_sfr}
\end{figure*}

Given the validity of vertical dynamical equilibrium and agreement of $\W$ with the simple weight estimator $\PDE$ (\autoref{eq:PDE}), the PRFM theory postulates that the yield $\Upsilon_{\rm tot}=\Ptot/\Ssfr$ (calibrated from simulations) can be used to predict $\Ssfr$ from $\PDE$, which is calculated from large-scale galactic properties in observations.
Summing up all pressure components, we obtain the total pressure support and the corresponding feedback yield. We find median $\Upsilon_{\rm tot}=1500\kms$ for model {\tt R8-4pc} and $720\kms$ for model {\tt LGR4-2pc}, respectively.

As shown in \autoref{fig:P_sfr}, the new simulation results are overall in good agreement with the TIGRESS-classic suite for the relation between $\Ssfr$ and pressure or weight. In each panel, we directly compare our results with the fitting results from \citet{2022ApJ...936..137O} for measured midplane pressure, measured weight, and estimated weight:
\begin{subequations}
\begin{eqnarray}
\label{eq:SFR_P}
\log(\Ssfr)  &=&  1.18~\log(\Ptottwo/k_B) -7.43 \\
\label{eq:SFR_W}
\log(\Ssfr)  &=&  1.17~\log(\W/k_B) -7.32\\
\label{eq:SFR_PDE}
\log(\Ssfr)  &=&  1.21~\log(\PDEtwo/k_B) - 7.66.
\end{eqnarray}
\end{subequations}
We refrain from delivering a new fitting formula or making additional quantitative adjustments to the feedback yields given the limited parameter space covered in the present work. In the future, we will extend our parameter space survey, especially toward low metallicity regimes, to generalize the numerical calibration of feedback yield in the PRFM theory.

\section{Summary \& Discussion}\label{sec:summary_and_discussion}
\begin{figure}
    \centering
    \includegraphics[width=\linewidth]{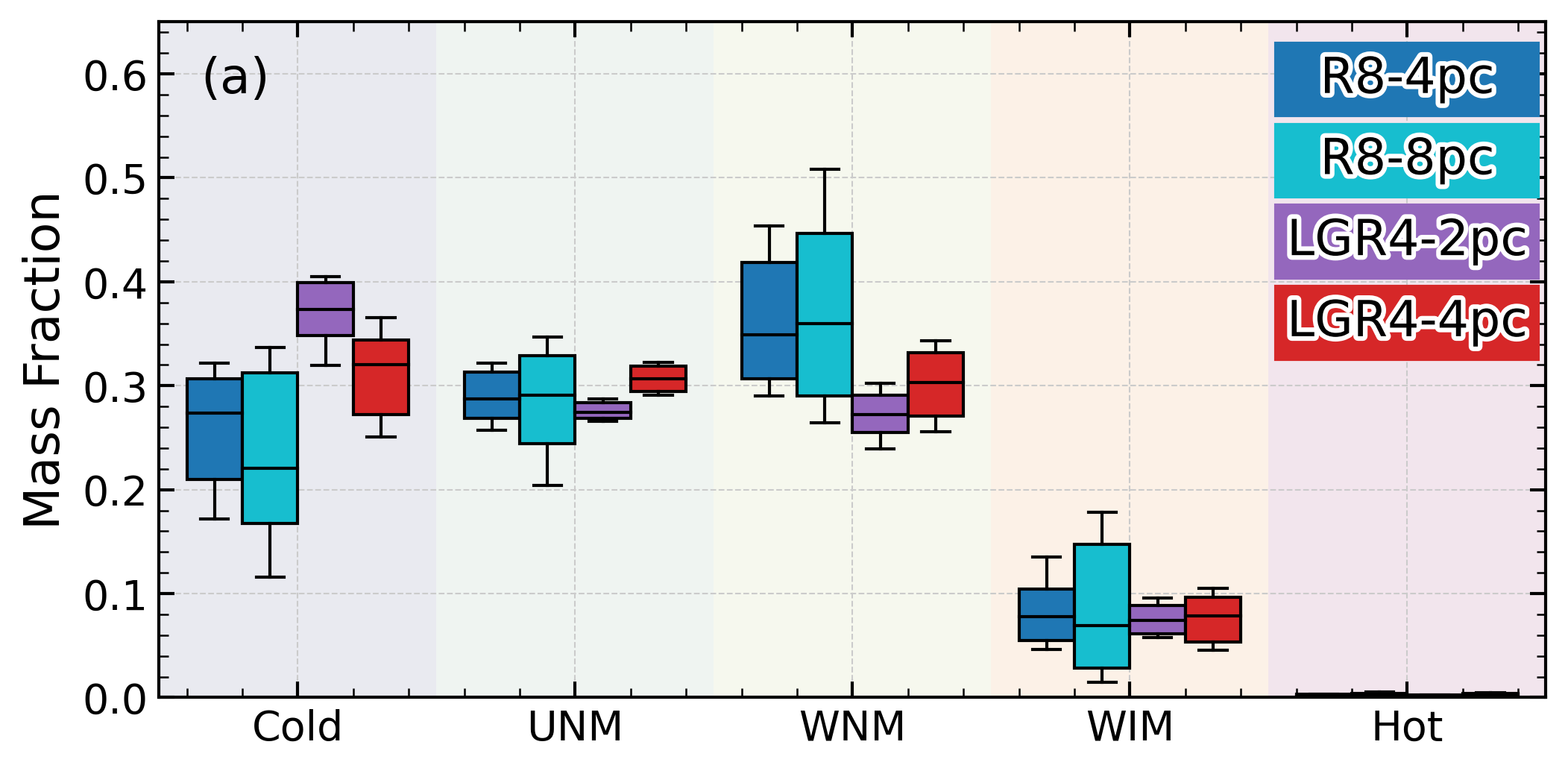}
    \includegraphics[width=\linewidth]{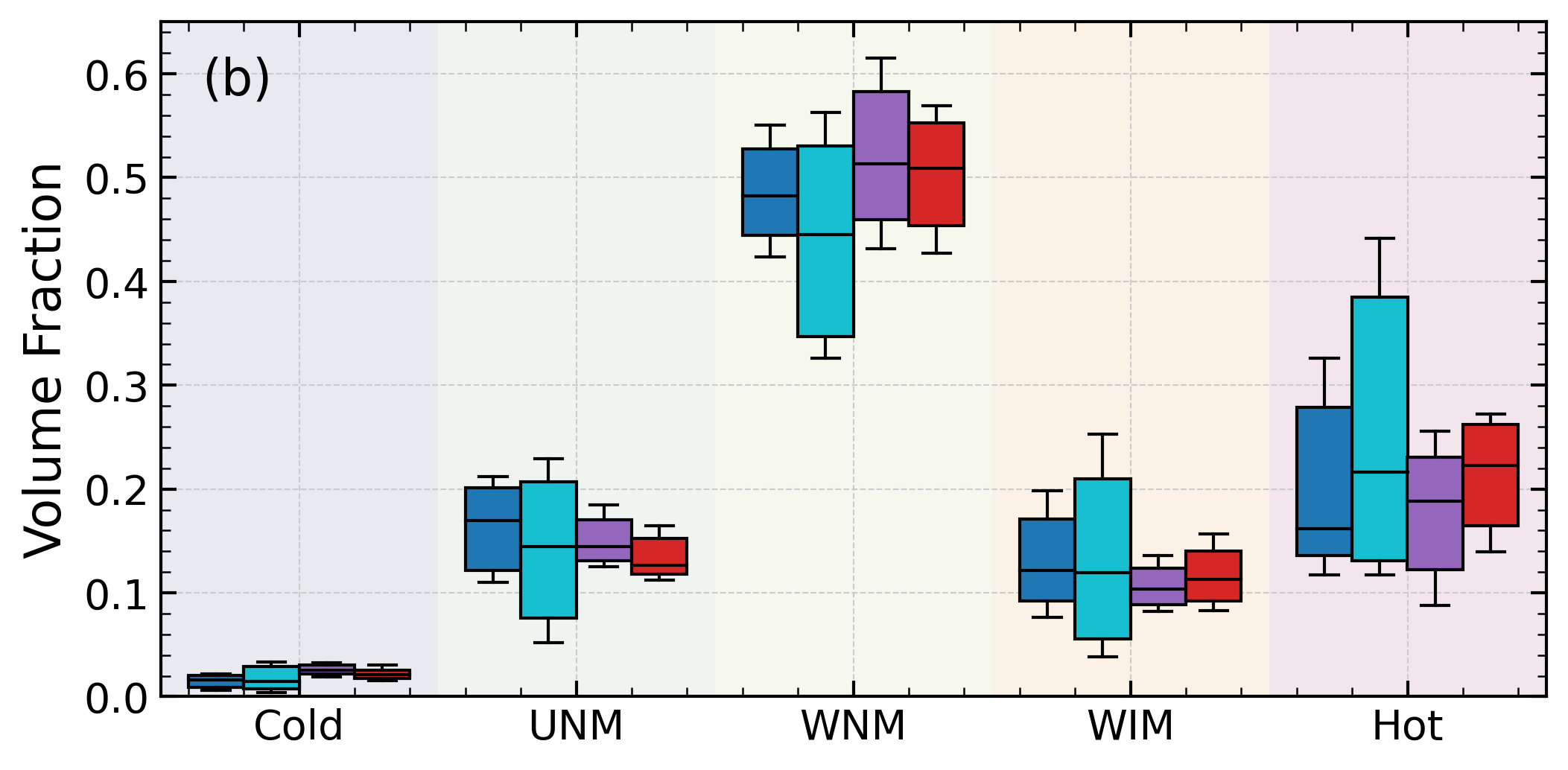}
    \includegraphics[width=\linewidth]{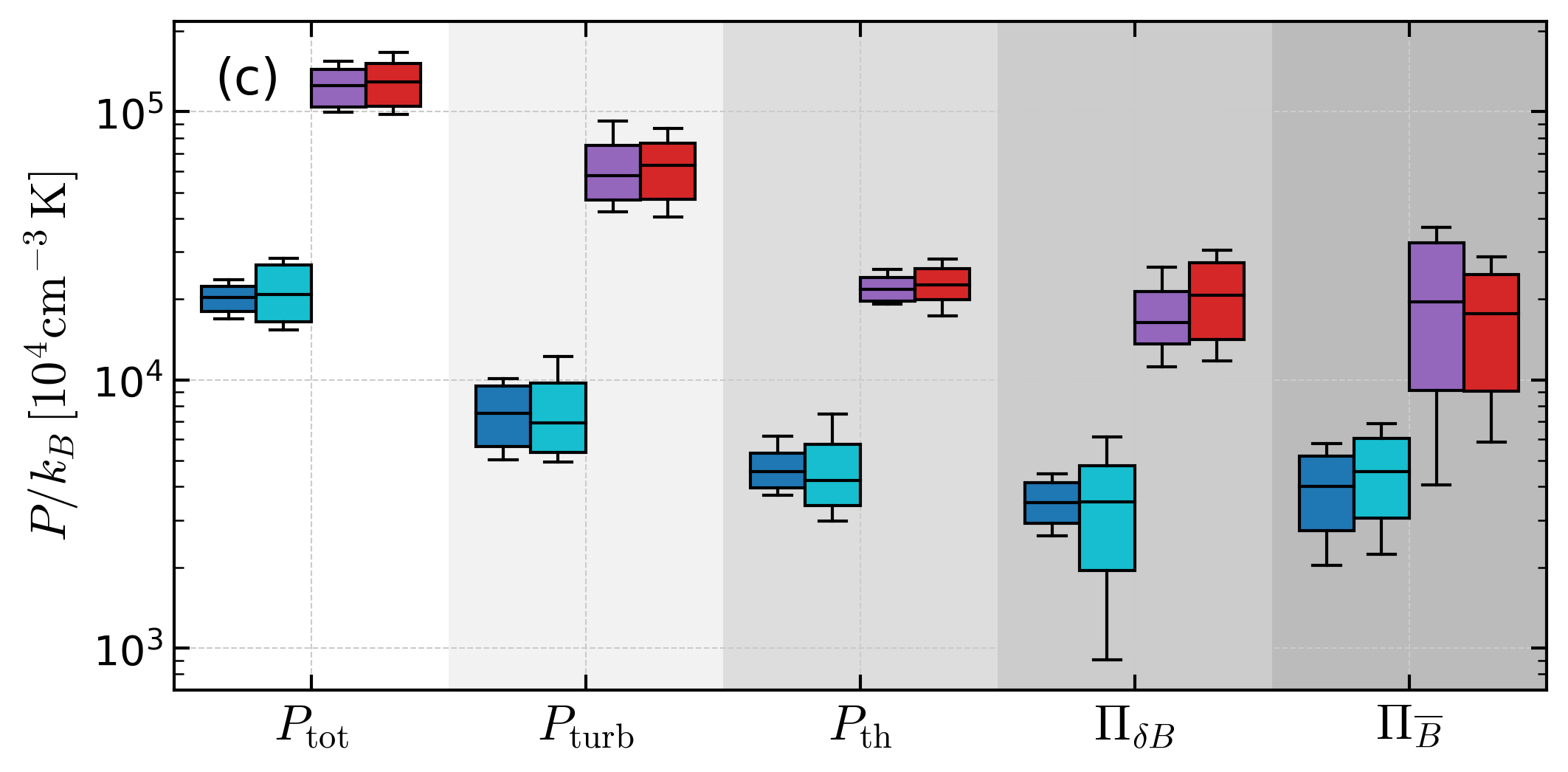}
    \includegraphics[width=\linewidth]{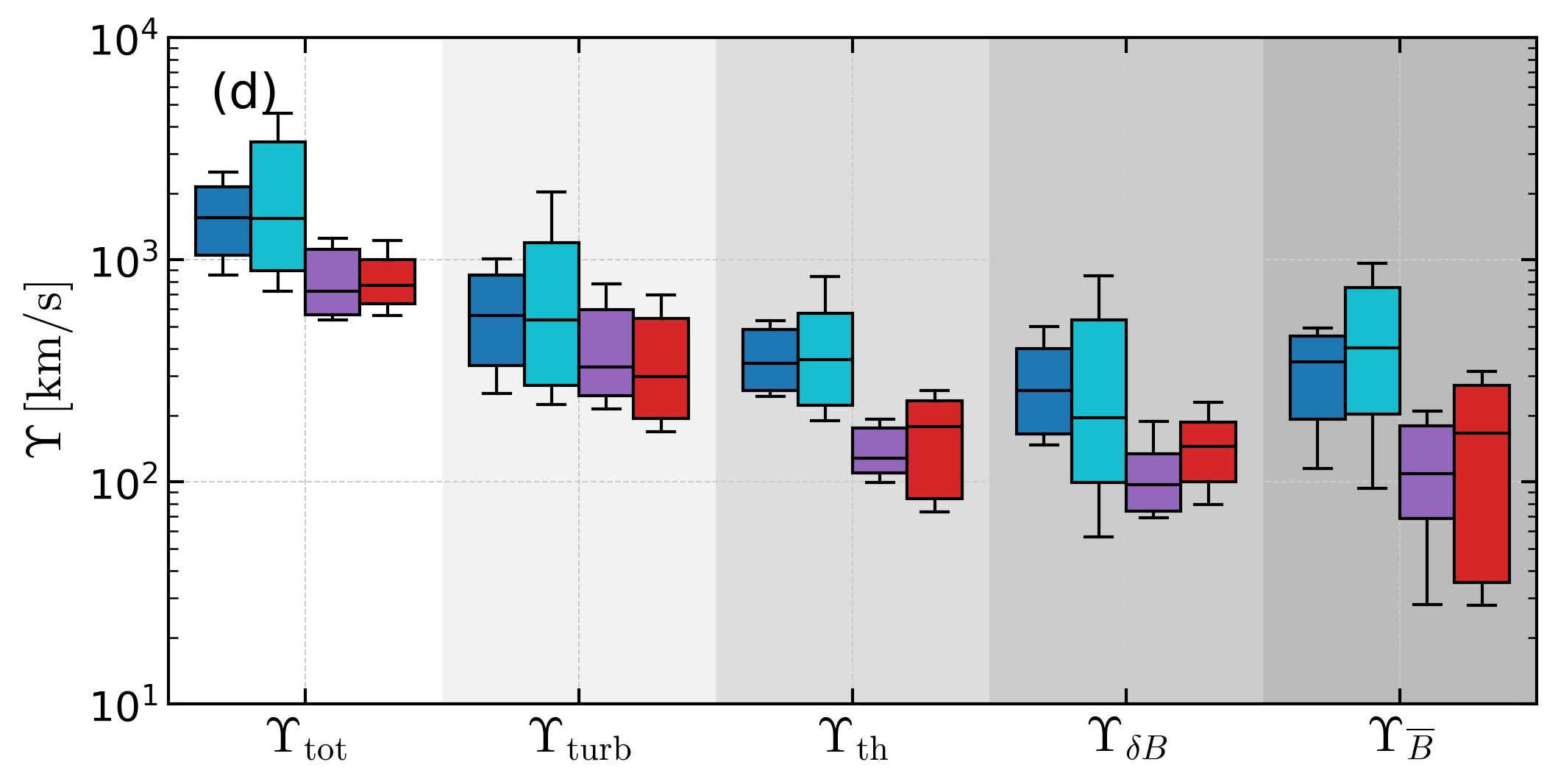}
    \caption{Summary of the main measured quantities. (a) Mass fraction and (b) volume fraction of each phase within $|z|<300\pc$. (c) Pressure components. (d) Feedback yields. The box and whisker enclose from 25$^{\rm th}$ to 75$^{\rm th}$ and from 16$^{\rm th}$ to 84$^{\rm th}$ percentiles, respectively, of the time evolution over $t\in(250,450)\Myr$ for {\tt R8} and $(250,350)\Myr$ for {\tt LGR4}. The median is shown as horizontal line within the box.}
    \label{fig:summary}
\end{figure}

\subsection{Summary of simulation results}\label{sec:summary}
We present first results from a new numerical framework that synthesizes the TIGRESS-classic computational model of the star-forming ISM \citep{2017ApJ...846..133K} with our recently developed non-equilibrium cooling and radiation (NCR) module \citep{KGKO}.
The detailed photochemical treatment and the effects of UV radiation from massive young stars are combined with the gravitational collapse/star formation and SN injection schemes implemented and tested in the TIGRESS-classic framework, in order to study the multiphase, turbulent, magnetized ISM self-consistently.

This paper considers two galactic conditions, one representing the solar neighborhood ({\tt R8}) and the other a higher density/pressure environment ({\tt LGR4}; close to the molecular gas weighted mean conditions in the PHANGS survey). We delineate the ISM properties, with a focus on the multiphase ISM distribution near the midplane (within a scale height). We then repeat the basic analysis done in \citet{2022ApJ...936..137O} to test, validate, and calibrate the PRFM star formation theory. The key measured quantities from our analysis are summarized in \autoref{tbl:satprop}, \autoref{tbl:pressure}, and \autoref{fig:summary}.

We summarize the ISM phase distributions by mass and volume within $|z|<300\pc$ in \autoref{fig:summary}(a) and (b). Near the galactic midplane (within one scale height of the disk), the cold, unstable, and warm neutral medium (\CNM{}, \UNM{}, and \WNM{}) occupies about 25\%, 30\%, and 35\% by mass and 2\%, 20\%, and 50\% by volume, respectively, in the solar neighborhood model {\tt R8}. The warm ionized medium (\WIM=\WPIM{}+\WCIM{}) contributes 8\% and 10\% by mass and volume, respectively, while the hot medium (\hot=\WHIM{}+\HIM{}) occupies about 15\% of the volume with a negligible mass contribution. It is important to keep in mind that there are large amplitude temporal fluctuations (up to a factor 2) in these values, as indicated by the box (25-75 percentiles) and whisker (16-84 percentiles) in \autoref{fig:summary}. Moving to conditions of higher gas surface density, total pressure, and SFR with model {\tt LGR4}, the mass contribution from \REV{colder components (i.e., \CMM{} and \CNM{})}{the \Cold{} (=\CMM{}+\CNM{}) component} increases, while the volume filling factors remain more or less \REV{constant}{the same}. For both models, the mass fractions of \Cold{} increase with higher resolution at the expense of \UNM{} and \WNM{}, although the change in {\tt R8} is well within the time fluctuation level. The volume fractions are converged up to the temporal variation level.

\autoref{fig:summary}(c) and (d) show the midplane pressure components and feedback yields for both models.
The two different resolutions give converged results for both {\tt R8} and {\tt LGR4}. The turbulent feedback yields are similar for {\tt R8} and {\tt LGR4}, with a slightly decreasing trend toward higher pressure environment.  The thermal feedback yield decreases as expected from {\tt R8} to {\tt LGR4}, due to higher shielding of FUV radiation field in higher density environments. In an upcoming paper (N. Linzer et. al. in prep.), we will analyze the radiation field in depth to validate and calibrate the global attenuation model used in TIGRESS-classic (see \autoref{eq:ftau}).
The magnetic feedback yields in {\tt R8} are generally larger than those of {\tt LGR4}; understanding the magnetic feedback yields requires further investigation of the galactic dynamo process, which in itself is a large and challenging area of research.  As shown in \autoref{fig:P_sfr}, the total feedback yields are quite similar to those reported in \citet{2022ApJ...936..137O}.  For {\tt R8}, $\Upsilon_{\rm tot}=1500\kms$, and for {\tt LGR4}, $\Upsilon_{\rm tot}=720\kms$.
Both models have similar $\sigma_{\rm eff,\twop} \approx 12-13 \kms$ and $\sigma_{\rm z, turb, \twop}\sim 7-8\kms$.

Finally, it is worth noting that the decrease in the \WNM{} mass fraction from model {\tt R8} to model {\tt LGR4} is at least qualitatively consistent with theoretical expectations \citep{2022ApJ...936..137O}.  The \WNM{} mass fraction may be written as
\begin{eqnarray}
f_{M,\WNM{}} &=&\frac{\rho_w}{\rho_{\rm tot}}f_{V,\WNM{}}\\ \nonumber
&=&\frac{P_{\rm th}}{P_{\rm tot}}\frac{\sigma_{\rm eff}^2}{c_w^2}f_{V,\WNM{}}
\end{eqnarray}
for $f_{V,\WNM{}}$ the volume filling factor, where we have used $P_{\rm th} = \rho_w c_w^2$ for $c_w$ the warm-gas sound speed (which is insensitive to galactic environment), $\rho_w$ for the typical density in the warm medium near the midplane, and $P_{\rm tot} \equiv \rho_{\rm tot} \sigma_{\rm eff}^2$ for $\rho_{\rm tot}$ the total midplane density.  From \autoref{tbl:satprop} and \autoref{tbl:filling}, $\sigma_{\rm eff}$ and $f_{V,\WNM}$ are very similar between model {\tt R8} and {\tt LGR4}, whereas \autoref{tbl:pressure} gives $\sim 30\%$ lower ratio of thermal to total pressure for {\tt LGR4} than that for {\tt R8}. About 30\% decrease in the \WNM{} mass fraction (\autoref{tbl:filling} and \autoref{fig:summary}) is consistent with the expectation.

\subsection{ISM Phase Balance and Distribution}\label{sec:disscuss_phase}

\subsubsection{Comparison to Milky Way Empirical Constraints on ISM Phases}\label{sec:galactic_ism}

Our phase distribution is overall in good agreement with multiwavelength galactic observations. \ion{H}{1} 21cm lines are the fundamental probe of the atomic ISM. An accurate determination of both gas column density and spin temperature requires \ion{H}{1} absorption line measurements paired with emission lines. Generally speaking, \WNM{} dominates 21~cm emission spectra, but \WNM{} is extremely faint in absorption due to its low density and high spin temperature (which can be as high as the gas temperature due to radiative excitation by Ly$\alpha$ resonant scattering; \citealt{1952AJ.....57R..31W,1959ApJ...129..536F,2020ApJS..250....9S}).
The detection of \WNM{} (and \UNM) in absorption requires highly sensitive absorption observations. There have been a number of sensitive absorption line surveys that determine mass fractions of different \ion{H}{1} components \citep{2003ApJ...586.1067H,2013MNRAS.436.2366R,2018ApJS..238...14M}. Using a simple radiative transfer model with multiple Gaussian components, a component detected in absorption with low or intermediate spin temperature ($T_s<250\Kel$ or $250\Kel<T_s<1000\Kel$) is considered to be \CNM{} or \UNM{}, while a component detected only in emission is \WNM{} (with a small fraction detected in absorption with large spin temperature). The mass contribution to total \ion{H}{1} column density of each component is roughly 30\%, 20-30\%, and 40-50\% for \CNM, \UNM, and \WNM, respectively, generally consistent among surveys.
From \autoref{tbl:filling}, the mass fractions of the cold, unstable, and warm neutral medium in model {\tt R8} are $\sim$ 0.3, 0.3, 0.4, generally consistent with current empirical constraints for \REV{the Milky Way}{the solar neighborhood} to the extent that they are available.

The observational measurement of thermal pressure using \ion{C}{1} fine structure lines shows a lognormal distribution with a mean at $P_{\rm th, \CNM}\sim 4000\Kel\pcc$ and an rms dispersion of 0.175 dex \citep{2011ApJ...734...65J}. We obtain the mass-weighted pressure PDF in {\tt R8} with mean and standard deviation of $\sim 1500\Kel\pcc$ and 0.27 dex for \CNM{}, $\sim 3700\Kel\pcc$ and 0.35 dex for \UNM{}, and $\sim 4000\Kel\pcc$ and 0.36 dex for \WNM{}.

Observations of H$\alpha$ and pulsar dispersion measures suggest that \WIM{} forms a thick layer with scale height of $\sim 1-2\kpc$ \citep{1989ApJ...339L..29R,1991IAUS..144...67R,1993ApJ...411..674T,2008ApJ...686..363H,2008PASA...25..184G}.
One can deduce the volume-averaged midplane electron density $\langle n_e \rangle \sim 0.02-0.05\pcc$ (by using dispersion measures to pulsars with known distances) and filling factor $f_{V, \WIM} \sim 0.05-0.15$ (by combining emission measures and dispersion measures) of the diffuse \WIM{} \citep{1987ASSL..134...87K,2001RvMP...73.1031F,2008PASA...25..184G}. For the midplane number density of total gas $\abrackets{\nH} \sim 0.5-1\pcc$ \citep[e.g.,][]{1990ApJ...365..544B,2015ApJ...814...13M}, the mass fraction of \WIM{} at the midplane is $\abrackets{n_e}/\abrackets{\nH}\simlt 10\%$. The mass fraction of \WIM{} of $f_{M, \WIM} \sim 7\%$ within $|z|<300\pc$ (see \autoref{tbl:filling}) is very much consistent with this empirical result.


Direct measurement of the hot gas in X-rays is difficult due to its low density. Also, significant diffuse soft X-ray emission is from the Local Bubble \citep{1987ARA&A..25..303C}. Soft X-ray radiation from larger scales is presumably absorbed; for example, the band averaged absorption cross section at $\sim 0.25$~keV is $\sim10^{-20} \cm^2{\rm H}^{-1}$ \citep{1990ApJ...354..211S}, yielding the mean free path $\sim 30 \pc (\nH/1\pcc)^{-1}$. Direct observational constraints on the larger scale distribution of likely pervasive hot gas in our Galaxy are still lacking.

\subsubsection{Comparisons to Self-consistent Numerical Models of the Star-Forming ISM}\label{sec:co_regulation_theory}

Because the SFR and the ISM thermal and dynamical state co-regulate each other, one cannot be considered independently of the other.
A theoretical model that explicitly addresses co-regulation,
computing the SFR  needed to maintain the thermal properties of the warm and cold ISM, was introduced by \citet{2010ApJ...721..975O}; this and subsequent theoretical developments are summarized in \citet{2022ApJ...936..137O}.

Several groups have recently developed numerical frameworks that solve (magneto)hydrodynamics with cooling and heating, including stellar feedback (of various forms) from star clusters that are self-consistently formed via gravitational collapse.

Our own numerical studies began with a focus on just the warm and cold ISM, with feedback in the form of momentum injection and heating, both proportional to $\Ssfr$ \citep{2011ApJ...743...25K,2013ApJ...776....1K,2015ApJ...815...67K}. These simulations, with a wide range of $\Sgas$, showed that a quasi-steady state is reached, validating vertical dynamical equilibrium.
For a solar neighborhood model (QA10 in \citealt{2013ApJ...776....1K}), the values of $\Ssfr\sim1.5\times10^{-3}\sfrunit$ and the midplane pressure (=weight) $\sim 10^4k_B\pcc\Kel$ were about a factor of two lower than those reported here (and from TIGRESS-classic) due to missing magnetic support and slightly weaker turbulence ($H\sim 80\pc$ vs. $220\pc$ and $\sigma_{z,{\rm turb}}\sim5\kms$ vs. $7-8\kms$). Coincidentally, the total feedback yield (without magnetic contribution) in \citet{2013ApJ...776....1K} is similar to that of the current simulations as the fixed $(p_*/m_*)=3\times10^3\kms$ adopted in the earlier work was higher than the effective $(p_*/m_*)_{\rm eff}\sim10^3\kms$ realized via self-consistent expansion of SNe driven bubbles \citep{2017ApJ...834...25K}.

\citet{2017ApJ...846..133K} introduced the TIGRESS-classic framework, with full treatment of the hot ISM. Direct comparison with the TIGRESS-classic suite results from \citet{2022ApJ...936..137O} regarding SFR, pressures, and feedback yields  show  overall consistent results with the current work, modulo slightly larger value of $\Upsilon_{\rm th}$ and hence $\Upsilon_{\rm tot}$ here (see \autoref{sec:prfm}). The lack of local shielding of FUV in  TIGRESS-classic tends to result in lower \Cold{} mass fraction ($f_{M, \CNM+\UNM}\sim 30\%$ in TIGRESS-classic vs. $f_{M,\Cold}\sim f_{M,\UNM}\sim30\%$ in TIGRESS-NCR). Inclusion of the ionizing radiation in TIGRESS-NCR converts significant \WNM{} into \WPIM{} ($f_{M,\WPIM}\sim7-8\%$), which is similar to the value obtained from the post-processing of TIGRESS-classic \citep{2020ApJ...897..143K}.

\citet{2014A&A...570A..81H} and \citet{2018A&A...620A..21C} are similar to our earlier work \citep{2013ApJ...776....1K,2015ApJ...815...67K} in terms of their SN feedback being mostly in the form of momentum injection without creating hot gas. The velocity dispersion in their models (which have $\Sgas=20\Surf$) is about $5-7\kms$, which is slightly lower than both of our new models. The magnetic fields tend to reduce SFR up to a factor of 2, with more reduction in the strong rotation case. Given that the magnetic feedback yield is about $30-40\%$ of the total, \citet{2015ApJ...815...67K} found similarly higher SFR in non-magnetized cases.  Comparisons between magnetized and unmagnetized cases using the TIGRESS-NCR framework will be investigated in a separate paper.

The SILCC framework first introduced by \citet{2015MNRAS.454..238W} focuses on  solar neighborhood ISM modeling, with particular emphasis on hydrodynamical evolution with a hydrogen and carbon chemistry network \citep[][collectively called {\sc SGChem}]{2007ApJS..169..239G,2007ApJ...659.1317G,2012MNRAS.421..116G}.
\citet[][]{2017MNRAS.466.1903G} added sink particle treatments and star formation via gravitational collapse in the SILCC framework.
They emphasized the role of stellar winds  shutting off further accretion after sink formation. They found the  resulting SFR surface density and ISM properties (mostly focused on hot gas filling factor) are  sensitive functions of the density threshold for sink particle formation. The highest density threshold model ($n_{\rm thresh}=10^4\pcc$) yields $\Ssfr\sim10^{-3}\sfrunit$, while the low density threshold model ($n_{\rm thresh}=10^2\pcc$) experiences a strong initial burst of star formation with $\Ssfr>10^{-2}\sfrunit$.
\citet{2017MNRAS.466.3293P} included radiation transfer for ionizing UV (without radiation pressure and with a constant FUV background), with the same treatment of SNe and stellar winds, and a high density threshold. Their models with and without ionizing radiation (with both SNe and stellar winds) show similar $\Ssfr\sim10^{-3}\sfrunit$ but the inclusion of UV radiation gives smaller $f_{V,h}$ of $\sim 20-30\%$, larger warm gas filling factor, and reduced ${\rm H}_2$ gas mass (about a factor two) at the end of their simulation ($\sim70\Myr$). However, these quantities were still evolving, and the short runtime of their simulations makes it unclear whether the reported values are representative values in the statistical steady state of these models.
The SFR obtained by \citet{2017MNRAS.466.3293P} in their simulations with SNe, stellar winds, and ionizing radiation is similar to what we obtain here for the {\tt R8} model, $\Ssfr=3\times 10^{-3} \sfrunit$.

Recently, \citet{2021MNRAS.504.1039R} conducted simulations using the SILCC framework with a more comprehensive feedback model including SNe, stellar winds, UV radiation, and CRs, as well as magnetic fields. By systematically turning on and off each feedback process, they found a progressive decrease in $\Ssfr$, $f_{V,h}$, and cold gas mass fraction with more feedback. The impact of CRs is not significant (given the short evolution time of $\sim 100\Myr$), and the model with SN, stellar winds, and radiation (called SWR) shows $\Ssfr\sim 1.5-2\times10^{-3}\sfrunit$, similar to what we find and  to observations. Within $|z|<250\pc$, their SWR model shows $f_{V,h}\sim 50\%$ ($35\%$ with CRs) and $f_{M, {\rm cold}}\sim 50\%$; both are larger than what we find here.
One potential reason is that their FUV radiation was assumed to be constant over time so that the thermal balance in the volume filling warm and cold ISM may not be fully self-consistent. EUV radiation was transferred using a tree-based backward ray tracing method \citep{2021MNRAS.505.3730W}, which is inherently less accurate than the direct ray tracing method we adopt here, especially behind regions of strong shielding (pervasive for EUV due to the huge cross section of neutral hydrogen against ionizing radiation). Finally, due to short evolution time ($t<100\Myr$), their measurements include an initial burst period ($25\Myr<t<100\Myr$), which may bias the hot gas filling factor toward higher values.

\citet{2021ApJ...920...44H} developed a local simulation that handles time-dependent hydrogen chemistry on-the-fly using a chemistry network based on {\sc SGChem}, and explored the effect of metallicity. Their radiation treatment is approximate: the (spatially-constant) unattenuated UV radiation field and CR ionization rate are scaled by recent star formation, with a local attenuation factor for FUV radiation applied using a tree-based method \citep{2012MNRAS.420..745C}. Photoionization is treated using an iterative Str\"omgren sphere approach \citep{2017MNRAS.471.2151H}. Although properties of the ISM phase structure from this simulation were not explicitly discussed, $\Ssfr\sim2-3\times10^{-3}\sfrunit$ and the mass fraction of warm ionized medium ($\sim5-10\%$) for the solar metallicity model are consistent with observations and our results.

\subsection{Future Perspectives}\label{sec:future}

The new simulation framework, TIGRESS-NCR, presented in this paper provides a promising tool for modeling the star-forming ISM. The main advance from the TIGRESS-classic framework is including direct UV radiation transfer and explicit chemical abundance calculations.  These extensions allow us to examine more detailed aspects of ISM physics, and enable us to explore new parameter space beyond the conditions that apply in normal, low-redshift spiral galaxies like the Milky Way. One immediate application is to explore low metallicity environments that are common in local dwarfs and prevalent in all galaxies at high redshifts. Effects of metallicity on species abundances and the CO-to-H$_2$ conversion factor have been studied in previous work, with  \citet{2018ApJ...858...16G,2020ApJ...903..142G} post-processing the TIGRESS-classic suite with six-ray radiation transfer and steady-state chemistry, and \citet{2021ApJ...920...44H,2022ApJ...931...28H} using a tree-based shielding column calculation with time-dependent hydrogen chemistry combined with  steady state carbon/oxygen chemistry. Given \REV{}{the} more accurate methods for UV radiation transfer implemented in the TIGRESS-NCR framework, it will be very interesting to make comparisons with these works employing approximate radiation transfer.
\REV{}{Also, the capability of modeling time-dependent H$_2$ chemistry 
will be an important tool 
in understanding observed chemical abundances \citep[see][for CH$^+$ and hence warm diffuse H$_2$ abundances]{2022arXiv220910196G}, although higher resolution than is possible in the present simulations may be needed for many applications.}

With a suite of simulations at varying metallicity,
we can extend the theoretical understanding of SFR/ISM co-regulation to the low-metallicity regime, where the thermal feedback yield (and therefore thermal pressure) to become larger than other components because radiation easily propagates over large distances. This extension of the PRFM theory will be critical in developing a subgrid star formation recipe for large scale cosmological simulations.
Applying the TIGRESS-NCR framework to study regions with strong spiral structures will be
straightforward, since the TIGRESS-classic framework has already been successfully used for models of this kind \citep{2020ApJ...898...35K}.

Although TIGRESS-NCR represents a significant advance in resolving and modeling key physical processes, there is still more to be done.
First, we do not explicitly model CR transport.
Currently, we only include ionization and heating by low-energy CRs, with the
background value scaled with $\Ssfr$ and $\Sgas$ and attenuated in high density
environments (see \ref{sec:rt}). This is a physically and empirically motivated
prescription but lacks quantitative calibration from direct numerical modeling
and ignores the dynamical effect of CRs.
Full CR transport should include advection by the gas, streaming along magnetic
field lines at the (ion) Alf\'ven speed, and diffusion by scattering off of MHD
waves that are likely self-generated for GeV and lower energies
\citep{2021ApJ...922...11A,2022ApJ...929..170A}. TIGRESS-NCR provides a unique
laboratory for CR transport modeling as our framework produces a turbulent,
multiphase ISM with realistic magnetic field and ionization structure as well
as realistic, high-velocity hot galactic winds. Although $\sim$GeV CRs dominate
the total energy budget and are expected to be dynamically important
\citep{2018MNRAS.479.3042G}, low-energy CRs are responsible
for ionization in most of the ISM's mass. Therefore, spectrally resolved CR
transport is necessary \citep[e.g.,][]{2020MNRAS.491..993G,2022MNRAS.510.3917G}.

Thermal conduction, which is not included in our current framework, can alter
the hot gas properties. The conductive heat transport from hot gas (created by
SNe) to the warm/cold ISM leads to evaporation, although conductivity may be
suppressed perpendicular to the magnetic field \citep{1965RvPP....1..205B}. To
the extent that it can act, conductive evaporation maintains the hot gas
pressure while increasing its mass and decreasing its temperature
\citep[e.g.,][]{2019MNRAS.490.1961E}.  Conduction could certainly alter the
observable properties of diffuse X-ray emission from the hot gas, and
potentially change the hot gas mass fraction and volume filling factor.

\acknowledgements

\REV{}{We are grateful to the referee for the timely and helpful report.}
C.-G.K. and E.C.O. were supported in part by NASA ATP grant No. NNX17AG26G.
The work of C.-G.K. was supported in part by NASA ATP grant No. 80NSSC22K0717.
J.-G.K. acknowledges support from the Lyman Spitzer, Jr. Postdoctoral
Fellowship at Princeton University and from the EACOA Fellowship awarded by the
East Asia Core Observatories Association.
M.G. acknowledges support from Paola Caselli and the Max Planck Institute for Extraterrestrial Physics.
Partial support was also provided by
grant No. 510940 from the Simons Foundation to E. C. Ostriker.

Resources supporting this work were provided in part by the NASA High-End Computing (HEC) Program through the NASA Advanced Supercomputing (NAS) Division at Ames Research Center and in part by the Princeton Institute for Computational Science and Engineering (PICSciE) and the Office of Information Technology's High Performance Computing Center.

\software{{\tt Athena} \citep{2008ApJS..178..137S,2009NewA...14..139S},
{\tt astropy} \citep{astropy:2013,astropy:2018,astropy:2022}, 
{\tt scipy} \citep{2020SciPy-NMeth},
{\tt numpy} \citep{vanderWalt2011}, 
{\tt IPython} \citep{Perez2007}, 
{\tt matplotlib} \citep{Hunter:2007},
{\tt xarray} \citep{hoyer2017xarray},
{\tt pandas} \citep{mckinney-proc-scipy-2010},
{\tt CMasher} \citep{CMasher},
{\tt adstex} (\url{https://github.com/yymao/adstex})
}
\clearpage

\appendix

\section{Technical Details on Ray Tracing}\label{sec:A-rt}

\subsection{Plane-parallel approximation for FUV radiation field at high-altitude regions}\label{sec:A-rt-planeparallel}

For PE and LW bands, we calculate the area-averaged intensity at $z = \zpp$ as a function of the cosine angle $\mucos = \khat \cdot \zhat$ (assuming azimuthal symmetry) as
\begin{align}
    \langle I \rangle (\mucos_i) & \equiv \frac{\int I(x,y,\zpp; \, \mucos_i) dx dy}{\int dx dy} \\
                        & = \frac{1}{L_x L_y} \sum_{\rm rays} \frac{\Delta L_{\rm ray}}{2\pi \alpha \Delta x \Delta \mucos} \,,
\end{align}
where $\alpha = \rho \kappa_{\rm d}$ is the dust absorption cross section per unit volume, $\Delta x$ is the cell size, $\Delta L_{\rm ray}$ is the amount by which the luminosity of the incoming photon packet is reduced as it passes through a cell at $z=\zpp$ \citep{2017ApJ...851...93K}, and the summation is taken over all rays passing through the layer $z=\zpp$ with $\mucos_i \le \mucos < \mucos_{i+1}$. We discretize the cosine angle as $\mucos_i = i \Delta \mucos$ with $\Delta \mucos = 2/N_{\mucos}$ and $i = 0, 1, \dots, N_{\mucos}-1$ and adopt $N_{\mucos} = 64$. Assuming plane-parallel geometry, the radiation energy density and flux in the vertical direction at $z > \zpp$ are given by
\begin{align}
\langle \mathcal{E}_{\rm rad,p\textnormal{-}p} \rangle & = (2\pi\Delta \mucos/c) \sum_i \langle I \rangle (\mucos_i) e^{- \Delta \tau / \mucos_i} \,, \\
\langle F_{z, {\rm rad,p\textnormal{-}p}} \rangle & = 2\pi\Delta \mucos \sum_i \langle I \rangle (\mucos_i) e^{- \Delta \tau / \mucos_i} \mucos_i \,.
\end{align}
where $\Delta \tau = \int_{\zpp}^z \langle \alpha \rangle dz$ is the (area-averaged) dust optical depth integrated from $\zpp$ to $z$. Similar calculations are done for $z < -\zpp$.

\subsection{Convergence with respect to RT parameters}\label{sec:A-rt-conv}
\begin{figure*}
    \centering
    \includegraphics[width=\linewidth]{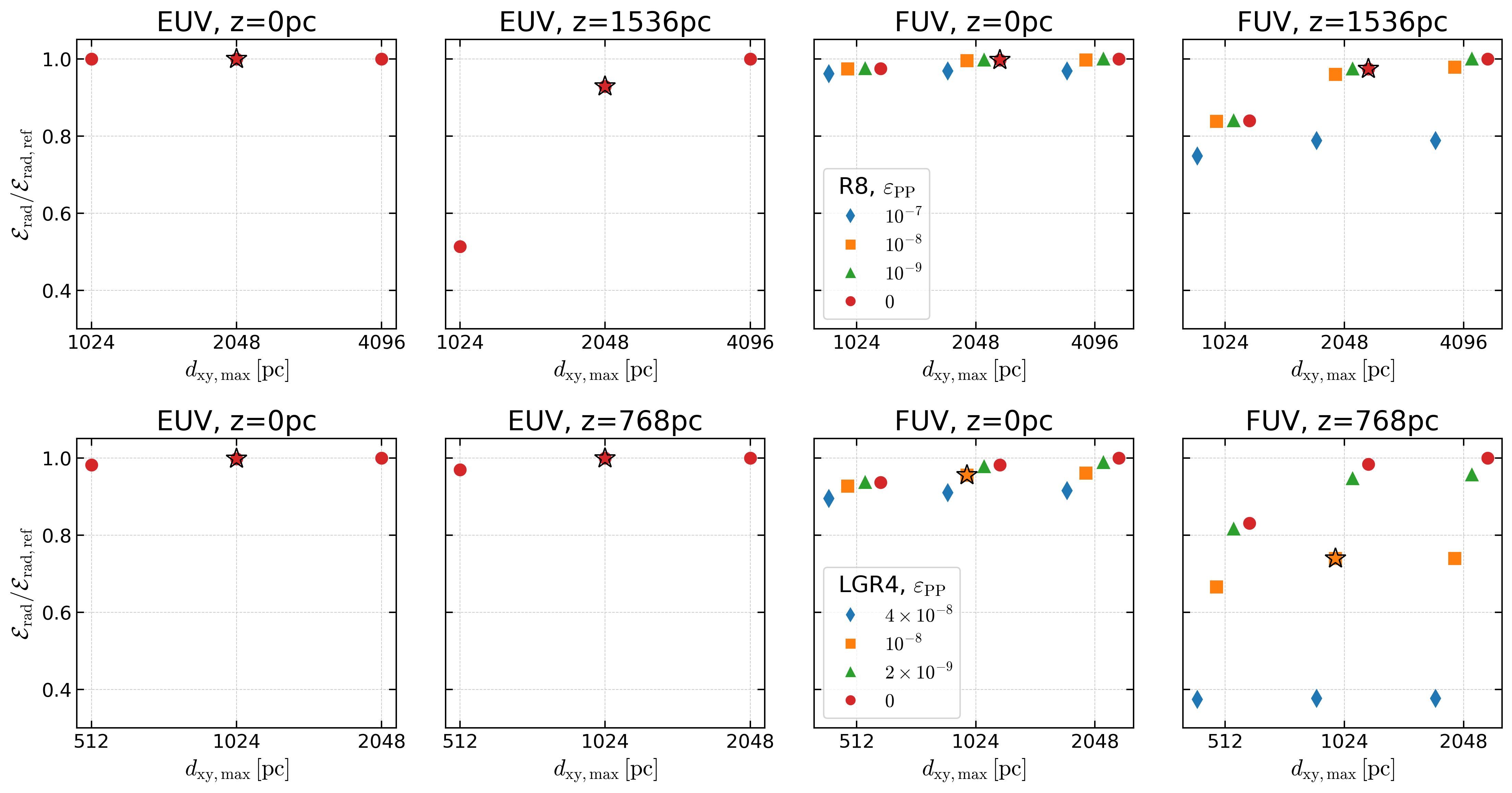}
    \caption{Convergence of the radiation energy density for {\tt R8} (top) and {\tt LGR4} (bottom). From left to right, we show the mean EUV (LyC) field at $z=0$ and $z=1536\pc$ and the mean FUV (PE+LW) field at $z=0$ and $z=768\pc$. Each point represents a model with different $\dxymax$ (x-axis) and $\epp$ (symbol; this condition is not applied to EUV). Our fiducial choice (shown as black star) is $\dxymax=2048\pc$ and $\epp=0$ for {\tt R8} and $\dxymax=1024\pc$ and $\epp=10^{-8}$ for {\tt LGR4}. }
    \label{fig:rt_conv1}
\end{figure*}

\begin{figure*}
    \centering
    \includegraphics[width=\linewidth]{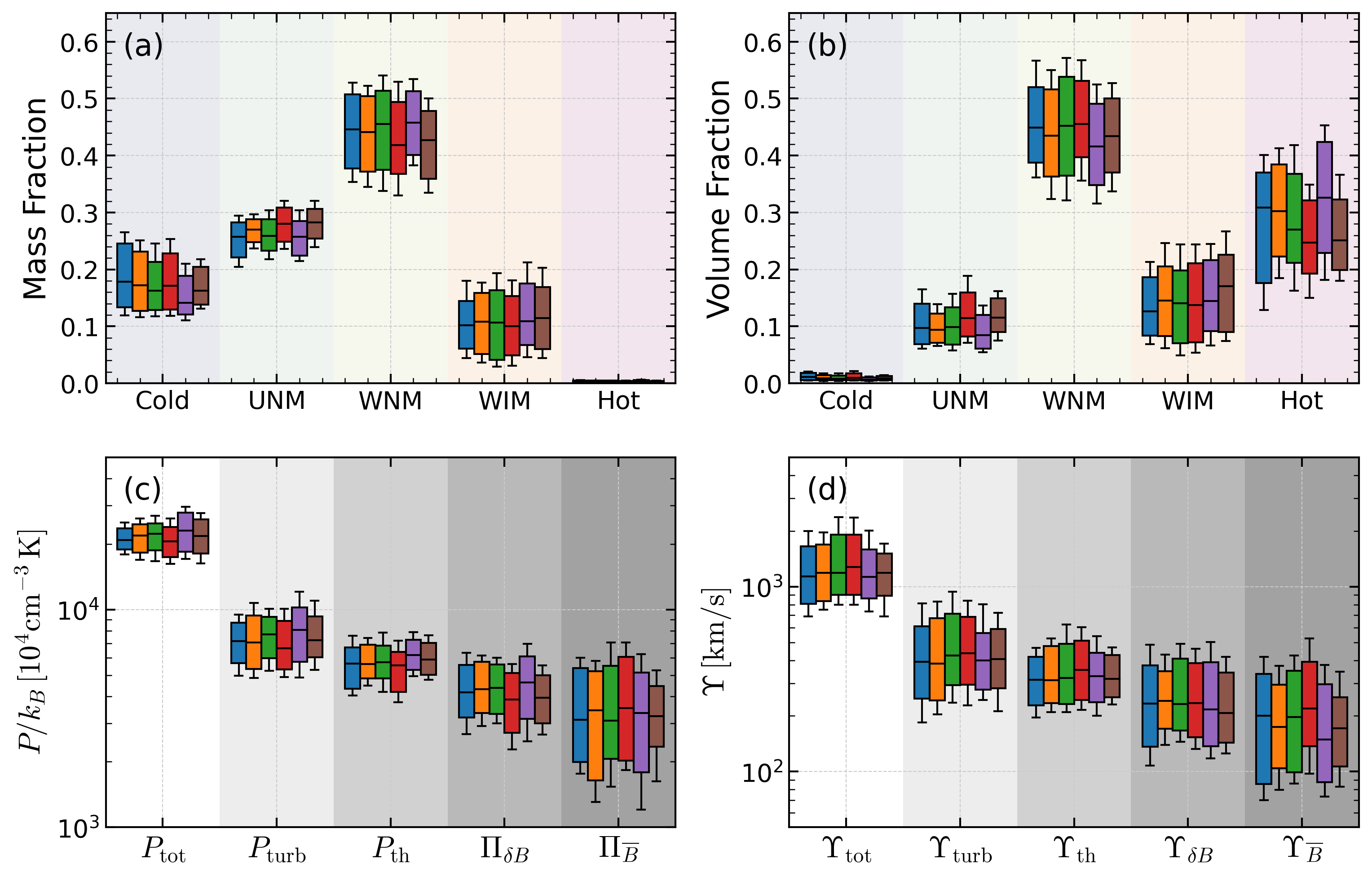}
    \caption{Convergence for {\tt R8} of (a) mass fraction, (b) volume fraction, (c) midplane pressure, and (d) feedback yield  with different ray-tracing numerical parameters $\dxymax$ and $\epp$. The box and whisker enclose 25$^{\rm th}$-75$^{\rm th}$ and 16$^{\rm th}$-84$^{\rm th}$ percentiles over $t\in(250,450)\Myr$, respectively. For each quantity, results from models A, B, C, D, E, and F are arranged left-to-right in increasing order of radiation transfer accuracy (see text for details). Our fiducial choice is $\dxymax=2048\pc$ and $\epp=0$ (purple).}
    \label{fig:rt_conv2}
\end{figure*}

We run two suites of simulations for different convergence tests. In the first suite, we take 10 snapshots (from 200~Myr to 300~Myr) of {\tt R8-8pc} and {\tt LGR4-4pc} and post-process the radiation fields, fixing MHD quantities and varying the radiation termination parameters ($\dxymax$ and $\epp$). Note that the FUV radiation field at $|z|>\zpp=300\pc$ is set by plane-parallel approximation (see \autoref{sec:A-rt-planeparallel}) rather than the direct radiation transfer. \autoref{fig:rt_conv1} shows the mean radiation energy densities of EUV and FUV at $z=0$ and $z=L_z/4$ with respect to the most accurate results, $(\dxymax, \epp)=(4096,0)$ for {\tt R8} and $(2048,0)$ for {\tt LGR4}. Both EUV and FUV near the midplane are well converged for our fiducial choice (shown as a black star) and generally better converged than those at high-$z$.
The FUV radiation field can be significantly underestimated if $\epp$ is large.
This implies that low luminosity sources (photon packets) make a significant contribution to the total FUV radiation field.

In the second suite, we restart simulations from Model {\tt R8-8pc} at 200~Myr with six choices of $(\dxymax, \epp)$; Model A: $(512,0)$, Model B: $(1024, 10^{-8})$, Model C: $(1024, 0)$, Model D: $(2048, 10^{-8})$, Model E: $(2048, 0)$, and Model F: $(4096, 0)$. Note that Model E is our fiducial choice (identical to {\tt R8-8pc}).
\autoref{fig:rt_conv2} shows the convergence of ISM mass and volume fractions within $|z|<300\pc$, midplane pressures, and feedback yields from this test suite. We present the range of values from the long-term evolution over $t\in(250,450)\Myr$. These quantities are well converged even in Model A with $\dxymax=512\pc$ and $\epp=0$ (blue, leftmost box).
Note that we did not repeat the same tests for model {\tt LGR4} because the additional expense was not merited, given the good convergence of the midplane radiation field shown in \autoref{fig:rt_conv1}.

We conclude that our fiducial choices result in good convergence in both radiation field properties and thermal/dynamical properties of the ISM. It is noteworthy that the accuracy of the radiation transfer can be reduced without affecting too much in the thermal and dynamical properties of the ISM (especially, those near the midplane). Unless the radiation field itself is the main subject of the study, our approach with early radiation termination can be used for a large parameter space exploration at a reduced cost.

\bibliography{ref,software}
\end{document}